\RequirePackage{lineno}
\documentclass[superscriptaddress,aps,preprint,amsmath,amssymb,floatfix]{revtex4}
\usepackage{graphicx,morefloats}
\usepackage{subfigure} 
\usepackage{color,soul}
\usepackage{caption}
\usepackage{bm}
\usepackage{times}
\usepackage[normalem]{ulem}

\usepackage{amsmath}
\usepackage{amssymb}
\usepackage{times}
\usepackage{hyperref}
\usepackage{bbm}

\def\ra{\rangle}
\def\la{\langle}
\def\up{\uparrow}
\def\dn{\downarrow}
\def\Hc{{\rm H.c.}}

\newcommand{\cepd}{Ce$_3$Bi$_4$Pd$_3$}

\newcommand{\qs}[1]{{\color{black} #1}}
\newcommand{\SBP}[1]{\color{black}{#1} \color{black}}

\begin{document}

\hyphenation{va-ni-sh-ing}

\begin{center}

\thispagestyle{empty}

{\large\bf Giant spontaneous Hall effect in a nonmagnetic\\ Weyl-Kondo semimetal}
\\[0.1cm]

S. Dzsaber$^1$, X.\ Yan$^1$, M.\ Taupin$^1$, G.\ Eguchi$^1$, A.\
Prokofiev$^1$, T.\ Shiroka$^{2,3}$,\\ P. Blaha$^4$, O.\ Rubel$^5$, S.\ E.\
Grefe$^6$, H.-H.\ Lai$^6$, Q.\ Si$^6$, and S.~Paschen$^{1,6,\ast}$\\[0.1cm]

$^1$Institute of Solid State Physics, Vienna University of Technology, 1040
Vienna, Austria\\

$^2$Laboratorium f\"ur Festk\"orperphysik, ETH Z\"urich, 8093 Zurich, Switzerland\\

$^3$Paul Scherrer Institut, 5232 Villigen PSI, Switzerland\\

$^4$Institute of Materials Chemistry, Vienna University of Technology, 1040 Vienna, Austria\\

$^5$Department of Materials Science and Engineering, McMaster University, 1280 Main Street West, Hamilton, Ontario, Canada L8S 4L8\\

$^6$Department of Physics and Astronomy, Rice Center for Quantum Materials, Rice University, Houston, Texas 77005, USA\\[-1.2cm]

\end{center}

\noindent
{\bf Nontrivial topology in condensed matter systems enriches quantum states of
matter, to go beyond either the classification into metals and insulators in
terms of conventional band theory or that of symmetry broken phases by Landau's
order parameter framework. So far, focus has been on weakly interacting systems,
and little is known about the limit of strong electron correlations. Heavy
fermion systems are a highly versatile platform to explore this regime. Here we
report the discovery of a giant spontaneous Hall effect in the Kondo semimetal
Ce$_3$Bi$_4$Pd$_3$ that is noncentrosymmetric but preserves time reversal
symmetry. We attribute \qs{this finding} to Weyl nodes---singularities of the
Berry curvature---that \qs{emerge in}
the immediate vicinity of the Fermi
level \qs{due to the Kondo interaction.} We stress that this phenomenon is distinct from the previously detected
anomalous Hall \SBP{effect}
in materials with broken time reversal symmetry;
instead, it manifests an extreme topological \SBP{response}
that requires a
beyond-perturbation-theory description of the previously proposed nonlinear Hall effect. \SBP{The large} magnitude
of the effect in even tiny electric and zero magnetic fields, as well as its
robust bulk nature may aid the exploitation in topological quantum
devices.}\vspace{0.6cm}

\noindent E-mail: $^{\ast}$paschen@ifp.tuwien.ac.at

\newpage

\noindent{\bf SIGNIFICANCE STATEMENT}\\
\noindent States of matter are traditionally classified by their symmetry, as exemplified by the distinction between a solid and a liquid. Topological quantum phases, on the other hand, are harder to characterize, and still harder to identify. This is especially so in electronic systems with strong correlations. In this work, we uncover a purely electric-field-driven ``giant'' Hall response---orders of magnitude above expectation---in one such material, and propose a mechanism how \SBP{it is driven by strong correlations.}
Our results will enable the identification of electronic topological states in a broad range of strongly correlated quantum materials, and may trigger efforts towards their exploitation in robust quantum electronics.

\vspace{1cm}

\newpage
\noindent{\bf INTRODUCTION}\\
Exploring effects of topology in weakly correlated condensed matter systems has
led to the identification of fundamentally new quantum phases and phenomena
\cite{NatPhys16.1a}, including the spin Hall effect \cite{Koe07.1}, protected
transport of helical fermions \cite{Hsi09.2}, topological superconductivity
\cite{Sas11.1}, and large nonlinear optical response \cite{Wu16.1,Ma17.1}. In
the recently discovered Weyl semimetals, bulk 3D Dirac cones describing massless
relativistic quasiparticles are stabilized by breaking either inversion symmetry
(IS) or time reversal symmetry (TRS) \cite{Arm18.1}. Key experiments in their
identification have been angle-resolved photoemission spectroscopy (ARPES)
\cite{Hua15.1,Xu15.2,Xu16.3} as well as magnetotransport measurements,
\SBP{providing evidence for} the chiral anomaly \cite{Hua15.2,Zha16.3,Arm18.1}---charge pumping
between a pair of Weyl nodes---via a large negative longitudinal
magnetoresistance or, for nanostructures in high magnetic fields, Weyl orbits
via quantum oscillation \cite{Mol16.1} or quantum Hall measurements \cite{Zha19.2}.

Whereas the perturbative effect of correlations on topological electronic states
is already under broad investigation \cite{Han19.1,Men19.2,Kan20.2,Sha20.1}, a
completely open question is how strong correlations drive either related or
entirely new topological states \cite{Yan14.2,Par16.1,Cas17.1,Ipp18.1,Rah19.1}.
To uncover them experimentally, not only new materials but also alternative
measurement techniques have to be found. For instance, to characterize the
recently proposed Weyl-Kondo semimetals \cite{Dzs17.1,Lai18.1}, neither of the
canonical probes for weakly interacting Weyl semimetals seems suitable: ARPES
experiments still lack the ultrahigh resolution needed to resolve strongly
renormalized bands, and magnetotransport signatures of the chiral anomaly or
Weyl orbits are expected to be suppressed by the reduced quasiparticle
velocities of strongly correlated materials \cite{Zha16.3}. Our discovery of a
giant spontaneous Hall effect in one such material not only identifies an ideal
new technique but also demonstrates that strong correlations can drive extreme
topological responses, which we expect to trigger much further work.

The material we have investigated is the noncentrosymmetric and nonsymmorphic heavy fermion semimetal Ce$_3$Bi$_4$Pd$_3$ \cite{Dzs17.1} that has recently been identified as a candidate
Weyl-Kondo semimetal \cite{Dzs17.1,Lai18.1}.  Its low-temperature specific heat
contains a giant electronic $c = \Gamma T^3$ term that was attributed to
electronic states with extremely flat linear dispersion \cite{Dzs17.1},
corresponding to a quasiparticle velocity $v^{\star}$ that is renormalized by a
factor of $10^3$ with respect to the Fermi velocity of a simple metal
\cite{Dzs17.1,Lai18.1}. This boosts the electronic $\Gamma T^3$ term to the
point that it even overshoots the Debye $\beta T^3$ term of acoustic phonons
\cite{Dzs17.1}. To scrutinize this interpretation by other, more direct probes
of topology is the motivation for the present work.
\newpage

\begin{figure}[t]
\begin{center}
\includegraphics[width=0.25\textwidth, width=1\textwidth]{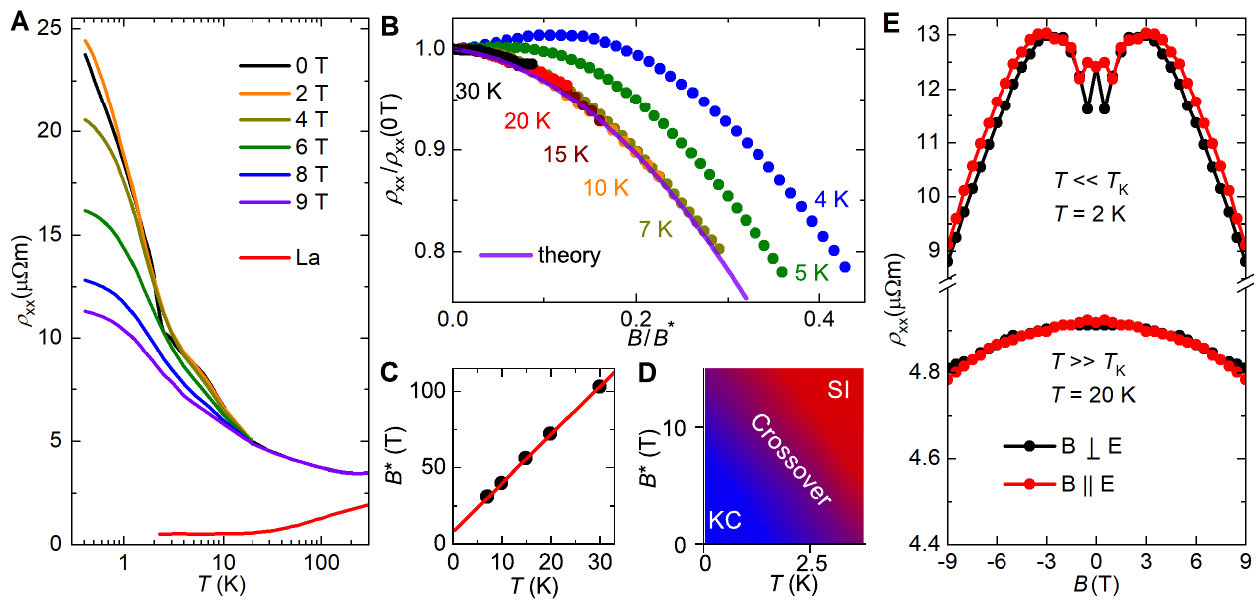}
\end{center}
\caption{\label{Fig1} {\bf Electrical resistivity and magnetoresistance of
Ce$_3$Bi$_4$Pd$_3$.} ({\bf A})~Temperature-dependent electrical resistivity
$\rho_{xx}$ of Ce$_3$Bi$_4$Pd$_3$ in various magnetic fields applied
perpendicular to the electric field ($B\perp E$, transverse magnetoresistance)
and of the nonmagnetic reference compound La$_3$Bi$_4$Pd$_3$ in zero field
(red). The feature in the 0\,T data of Ce$_3$Bi$_4$Pd$_3$ near 3\,K is due to the onset of the spontaneous Hall effect (Fig.\,\ref{Fig2}A), which leaves a finite imprint on $\rho_{xx}$ because of the giant Hall angle (Supplementary Information Sect.\,\ref{SI_RxxJump}) and because of slight contact misalignment (Supplementary Information Sect.\,\ref{misaligne}). ({\bf B}) Transverse magnetoresistance scaled to its zero-field value vs
scaled magnetic field $B/B^{\star}$, showing the collapses of data above 7\,K
onto the universal curve \cite{Sch83.1} (violet) expected for an $S = 1/2$ Kondo
impurity system in the incoherent regime, and a breakdown of the scaling for
temperatures below 7\,K (shown here by using $B^{\star}$ from the linear fit in
{\bf c},---also other choices of $B^{\star}$ cannot achieve scaling). ({\bf C})
Scaling field $B^{\star}$, as determined in ({\bf B}), vs temperature, showing a
linear-in-$T$ behavior as expected for a Kondo system in the single-impurity
regime. Fitting $B^{\star} = B^{\star}_0 (1 + T/T^{\star})$ to the data (red
straight line) yields $B^{\star}_0 = 10$\,T and $T^{\star} = 2.5$\,K. ({\bf D})~Temperature-field phase diagram displaying the single impurity (SI) and Kondo
coherent (KC) regime as derived in ({\bf C}). ({\bf E}) Transverse (black) and
longitudinal (red, $B||E$) magnetoresistance for temperatures well below (top)
and well above $T_{\rm K}$ (bottom). The data were symmetrized to remove any
spurious Hall resistivity contribution, and mirrored on the vertical axis for
clarity.}
\end{figure}

\noindent{\bf RESULTS}\\
We start by showing that Ce$_3$Bi$_4$Pd$_3$ is governed by the Kondo
interaction, and delineate the temperature and field range of Kondo coherence.
The zero-field resistivity of Ce$_3$Bi$_4$Pd$_3$ increases weakly with
decreasing temperature, whereas the nonmagnetic reference compound
La$_3$Bi$_4$Pd$_3$ is metallic (Fig.\,\ref{Fig1}A). This provides strong
evidence that the semimetallic character of Ce$_3$Bi$_4$Pd$_3$ is due to the
Kondo interaction. Below the single-ion Kondo temperature $T_{\rm{K}} = 13$\,K,
identified by associating the material's temperature-dependent entropy with a
spin $1/2$ ground state doublet of the Ce $4f^1$ wavefunction split by the Kondo
interaction \cite{Dzs17.1}, a broad shoulder in the resistivity at about 7\,K
signals the crossover to a Kondo coherent state (Fig.\,\ref{Fig1}A). As shown in
what follows, this is further supported by our magnetoresistance measurements
(Fig.\,\ref{Fig1}B,\,C). 


Transverse magnetoresistance isotherms (Fig.\,\ref{Fig1}B) in the incoherent
regime between 7 and 30\,K display the universal scaling typical of Kondo
systems \cite{Sin14.2,Map06.1}: $\rho_{xx}/\rho_{xx}(0\,{\rm T})$ vs
$B/B^{\star}$ curves all collapse onto the theoretically predicted curve for an
$S=1/2$ Kondo impurity system \cite{Sch83.1}, provided a suitable scaling field
$B^{\star}$ is chosen. The resulting $B^{\star}$ is linear in temperature
(Fig.\,\ref{Fig1}C). Fitting $B^{\star} = B^{\star}_0 (1 + T/T^{\star})$ to the
data (red straight line) yields $B^{\star}_0 = 10$\,T and $T^{\star} = 2.5$\,K,
which may be used as estimates of the field and temperature below which the
system is fully Kondo coherent (blue area in Fig.\,\ref{Fig1}D). Below 7\,K, the
scaling fails (Fig.\,\ref{Fig1}B), as expected when crossing over from the
incoherent to the Kondo coherent regime.

Before presenting our Hall effect results we show that, as anticipated, the
chiral anomaly cannot be resolved in the Kondo coherent regime. We find that, at
2\,K, the longitudinal and transverse magnetoresistance traces essentially
collapse (Fig.\,\ref{Fig1}E\,top). Because the amplitude $c_{a}$ of the chiral
anomaly,
being inversely proportional to the density of states \cite{Zha16.3},
is expected to scale as $c_{a} \propto (v^{\star})^{3}$, it is severely
suppressed by the strong correlations. The fact that, also at high temperatures, we do not observe signatures of the
chiral anomaly (Fig.\,\ref{Fig1}E\,bottom) \SBP{is consistent with our bandstructure calculations (to be presented later, Fig.\,\ref{Fig4}A), which reveal that the uncorrelated bandstructure contains Weyl nodes only far away ($> 100$\,meV) from the Fermi level.}

\begin{figure}[t!]
\begin{center}
\vspace{-1.1cm}

\includegraphics[width=1\textwidth]{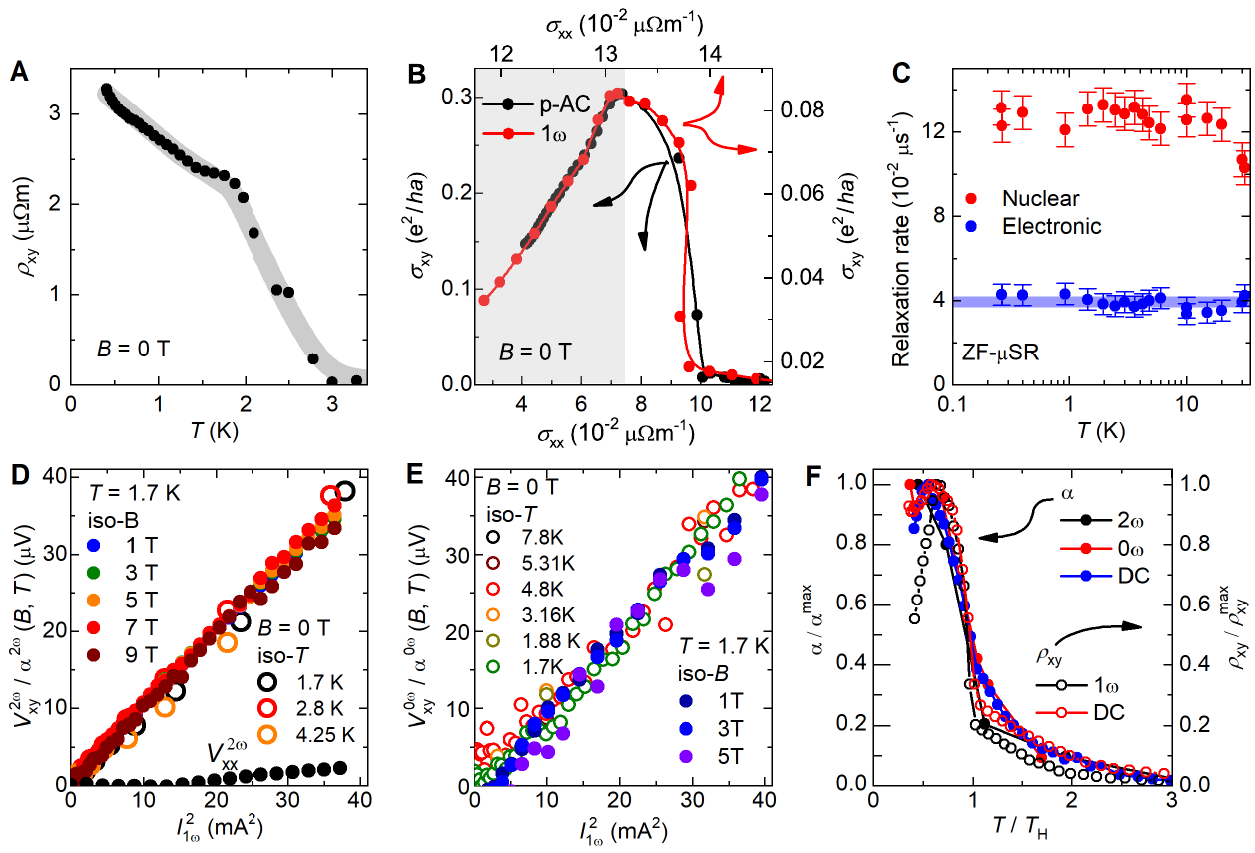}
\end{center}
\vspace{-0.7cm}

\caption{\label{Fig2} {\bf Spontaneous Hall effect of Ce$_3$Bi$_4$Pd$_3$.} ({\bf
A}) Temperature-dependent DC Hall resistivity $\rho_{xy}$ in zero external
magnetic field, showing a pronounced spontaneous Hall effect below 3\,K. Data
were taken without prior application of magnetic fields. ({\bf B}) Spontaneous DC
Hall conductivity $\sigma_{xy}$ in units of the 3D conductivity quantum vs
longitudinal conductivity $\sigma_{xx}$, with temperature as implicit parameter,
for the DC response of the sample in ({\bf A}) (bottom and left axes, black), and
for the $1\omega$ response in an AC experiment on a sample from a different
batch (top and right axes, red), both in zero magnetic field. In the Kondo
coherent regime (grey shading), $\sigma_{xy}$ is linear in $\sigma_{xx}$. ({\bf
C}) Temperature-dependent nuclear and electronic contributions to the muon spin
relaxation rate obtained from ZF-$\mu$SR measurements. The electronic
contribution is extremely small and temperature-independent within the error
bars, ruling out TRS breaking with state-of-the-art accuracy. Magnetization and specific heat measurements corroborate this finding (Supplementary Information Sects.\,\ref{SImagnetization} and \ref{SICp} with Figs.\,\ref{Fig:SI_hysteresis} and \ref{fig:Cp_S1}). ({\bf D}) Scaled $2\omega$ spontaneous Hall voltage vs square of
$1\omega$ driving electrical current in zero magnetic field for different
temperatures, and at 1.7\,K for various magnetic fields. ({\bf E}) Quantities
analogous to ({\bf D}) for the $0\omega$ spontaneous Hall voltage. ({\bf F}) Scaled
coefficients of square-in-current response $\alpha^{2\omega,0\omega,\rm{DC}}$
from panel ({\bf D}), ({\bf E}), and Fig.\,\ref{dc}, respectively (left axis), and
linear-in-current response $\rho_{xy}^{1\omega,\rm{DC}}$ from panel ({\bf B})
(right axis), as function of scaled temperature ($T_{\rm H}$ is the onset
temperature \SBP{of the}
spontaneous Hall signal). The absolute values of
$\alpha^{{\rm max},i}$, $\rho_{xy}^{{\rm max},i}$, and $T_{\rm H}$ are listed in
Table~\ref{SMtable}.}
\end{figure}

Our key observation\SBP{, presented next,} is a spontaneous (nonlinear, as will be discussed later) Hall effect which appears in
Ce$_3$Bi$_4$Pd$_3$ \SBP{as full Kondo coherence is established} below $T^{\star}$ (Fig.\,\ref{Fig2}A). The corresponding
spontaneous Hall conductivity $\sigma_{xy}$ reaches a considerable fraction of
the quantum of 3D conductivity (Fig.\,\ref{Fig2}B). The
experiment, using a pseudo-AC mode (Methods), was not only carried out in zero
external magnetic field, but also without any sample pre-magnetization process.
Hall contact misalignment contributions were corrected for \SBP{(Supplementary Information Sect.\,\ref{misaligne} and Figs.\,\ref{SMfig_ContactSketch}, \ref{SMfig2})}
and, thus, can also not account for the effect. Moreover, being in the Kondo
coherent regime, the local moments should be fully screened by the conduction
electrons. \SBP{The resulting paramagnetic state is evidenced by the absence of phase transition anomalies in magnetization and specific heat measurements (Supplementary Information Sects.\,\ref{SImagnetization}, \ref{SICp} and Figs.\,\ref{Fig:SI_hysteresis}, \ref{fig:Cp_S1}), as well as by}
state-of-the-art zero-field muon spin rotation ($\mu$SR) experiments. \SBP{The latter}
reveal an extremely small electronic relaxation rate that is
temperature-independent between 30\,K and 250\,mK (Fig.\,\ref{Fig2}C; for two
fully collapsing representative spectra, one well above and one well below
$T^{\star}$, see Supplementary Information Sect.\,\ref{SImuSR} and Fig.\,\ref{fig:ZF_LF_muSR}). This is unambiguous evidence that,
in the investigated temperature range and in particular across $T^{\star}$, TRS
is preserved in Ce$_3$Bi$_4$Pd$_3$. \SBP{In conjunction with the giant spontaneous Hall effect this} observation is striking---we are not aware of any other 3D material with preserved TRS that has shown a spontaneous Hall effect.

\begin{figure}[t]
\begin{center}
\includegraphics[width=1\textwidth]{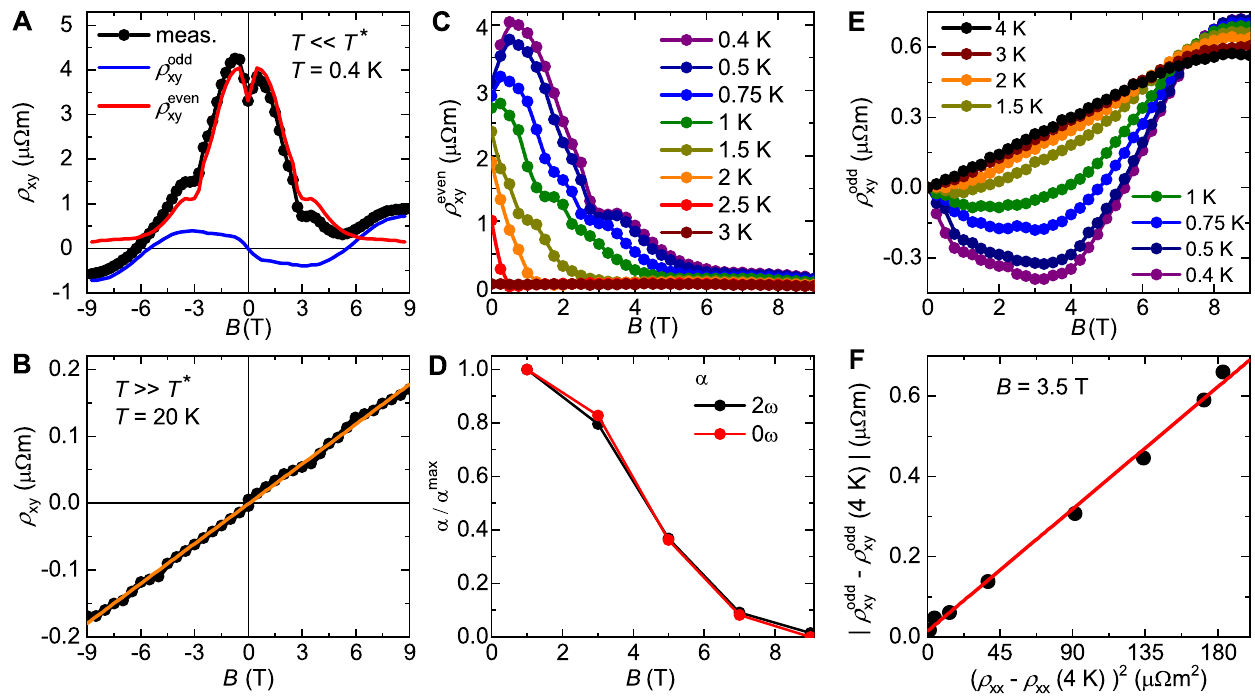}
\end{center}
\caption{\label{Fig3} {\bf Hall resistivity of Ce$_3$Bi$_4$Pd$_3$ in external
magnetic fields.} ({\bf A}) In the Kondo coherent regime below $T^{\star}$ and
$B^{\star}_0$ (Fig.\,\ref{Fig1}D), the magnetic field-dependent DC Hall
resistivity $\rho_{xy}(B)$ shows a pronounced anomalous Hall effect (AHE) and
can be decomposed into an odd-in-$B$ $\rho^{\rm{odd}}_{xy}(B)$ (blue) and an
even-in-$B$ $\rho^{\rm{even}}_{xy}(B)$ (red) component. ({\bf B}) Above
$T^{\star}$, $\rho_{xy}(B)$ is dominated by a linear-in-$B$ normal Hall effect.
({\bf c}) $\rho^{\rm{even}}_{xy}(B,T)$ is suppressed for $T > T^{\star}$ and $B
> B^{\star}_0$. ({\bf D})~Scaled coefficients of the $2\omega$ and $0\omega$
Hall voltage in an AC experiment (from Fig.\,\ref{Fig2}D,\,F) as function of
magnetic field. ({\bf E}) Below $T^{\star}$ and $B^{\star}_0$,
$\rho^{\rm{odd}}_{xy}(B,T)$ shows a pronounced AHE on top of a linear background
from the normal Hall effect. ({\bf F}) Amplitude of the odd-in-$B$ AHE,
estimated as the total odd-in-$B$ component at 3.5\,T (location of extremum)
minus its value at 4\,K, where the effect has disappeared (see ({\bf E})), vs
the square of the corresponding magnetoresistance difference
[$\rho_{xx}(T)-\rho_{xx}($4\,K)] at 3.5\,T, with $T$ as implicit parameter. The
observed quadratic dependence (red straight line) is in remarkable agreement
with expectations for the AHE due to broken TRS as $B$ is applied.}
\end{figure}
\begin{figure}[h!]
\begin{center}
\vspace{-1cm}

\hspace{-0.17cm}\includegraphics[width=0.45\textwidth]{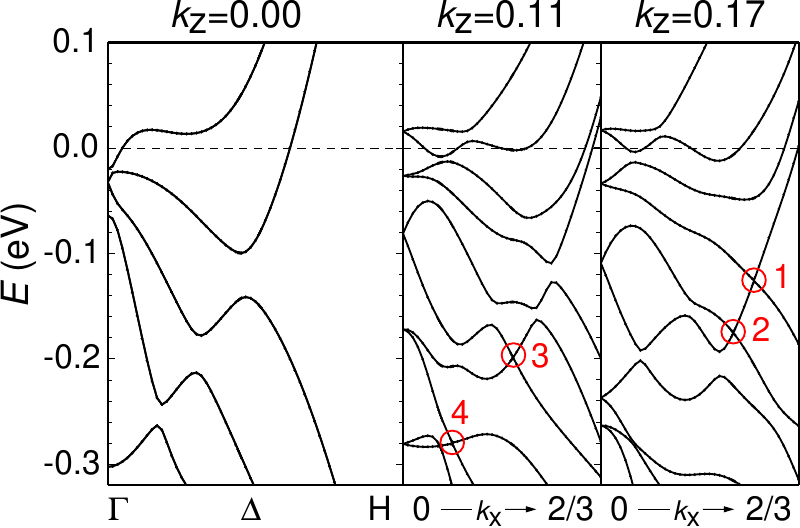}\\[0.2cm]
\includegraphics[width=0.442\textwidth]{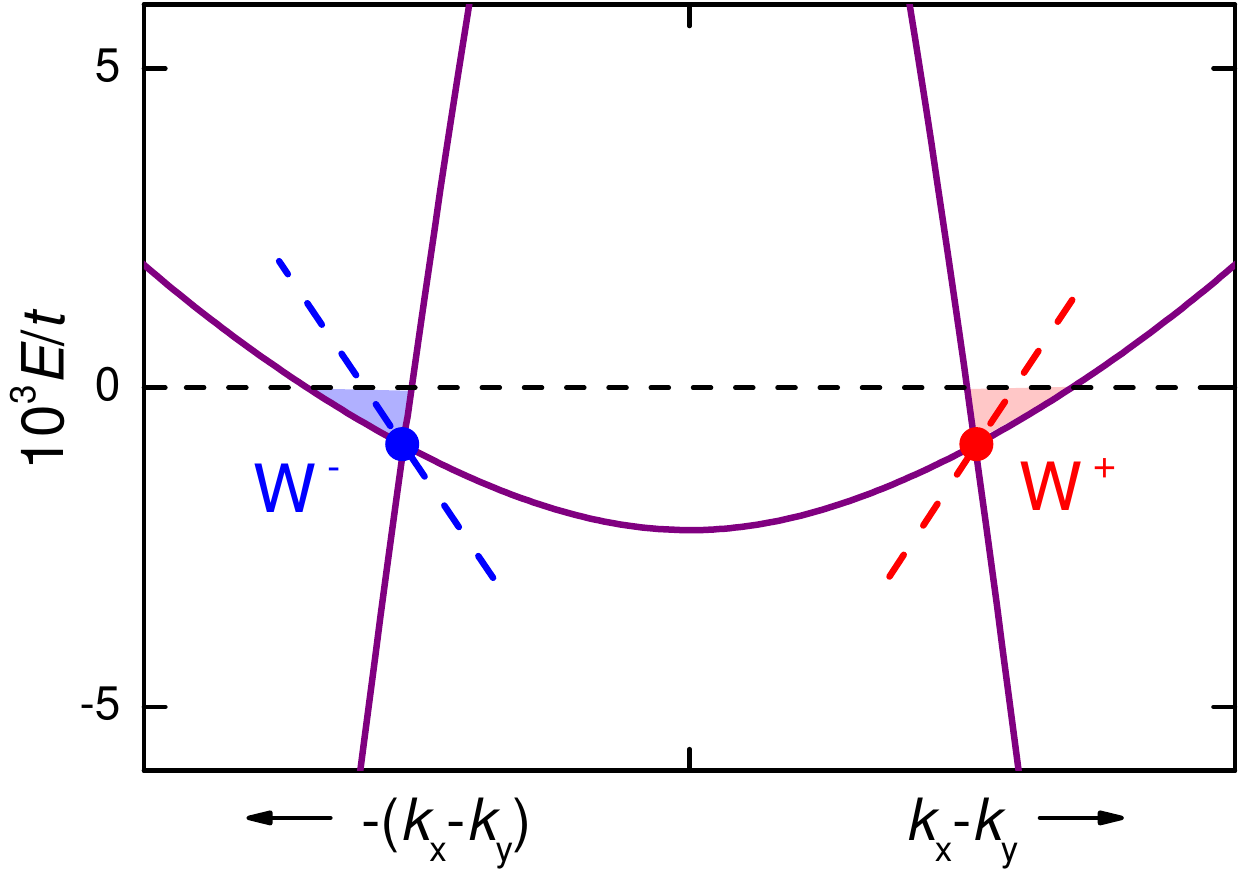}\\[0.2cm]
\includegraphics[height=0.19\textheight]{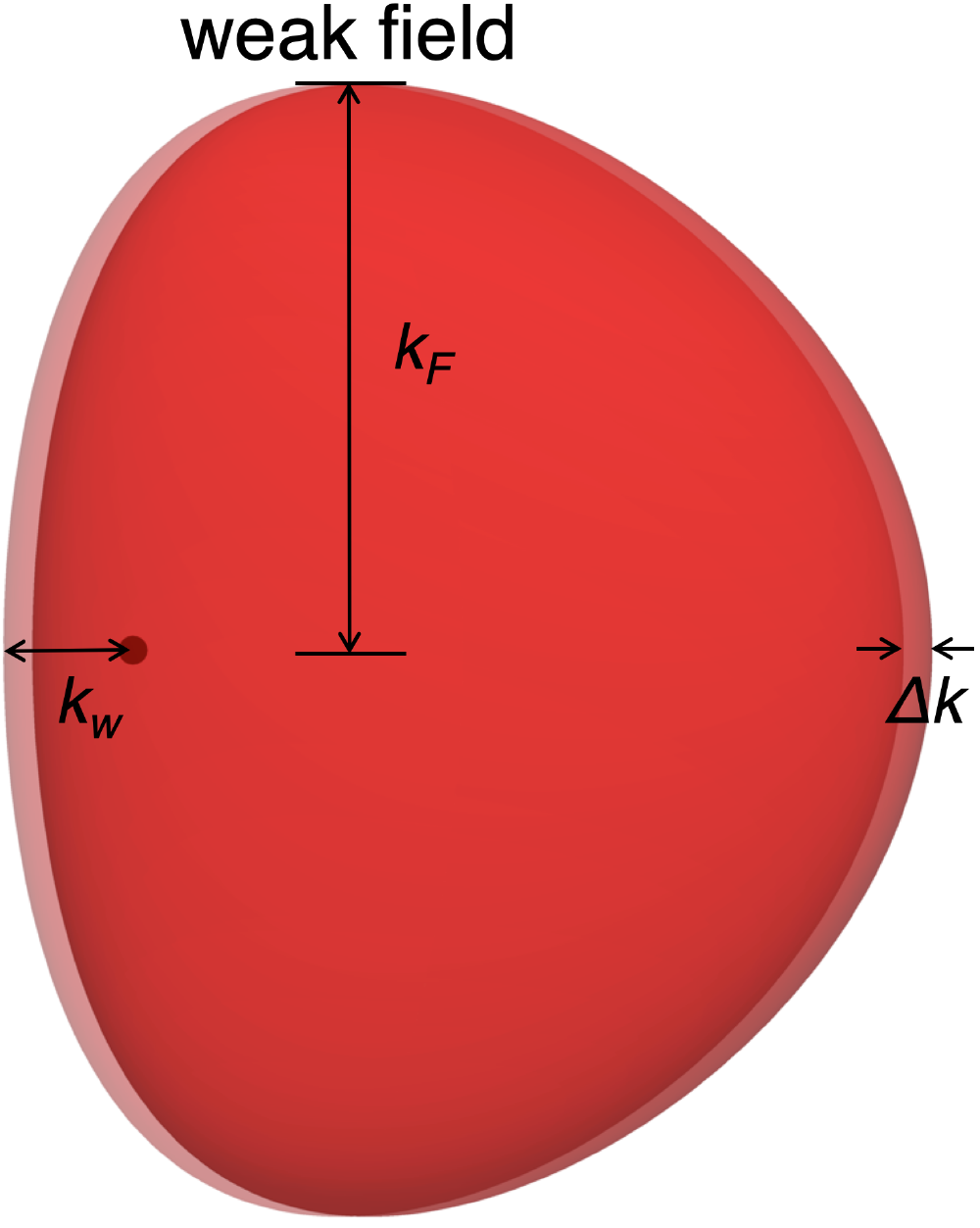}\hspace{0.3cm}\includegraphics[height=0.19\textheight]{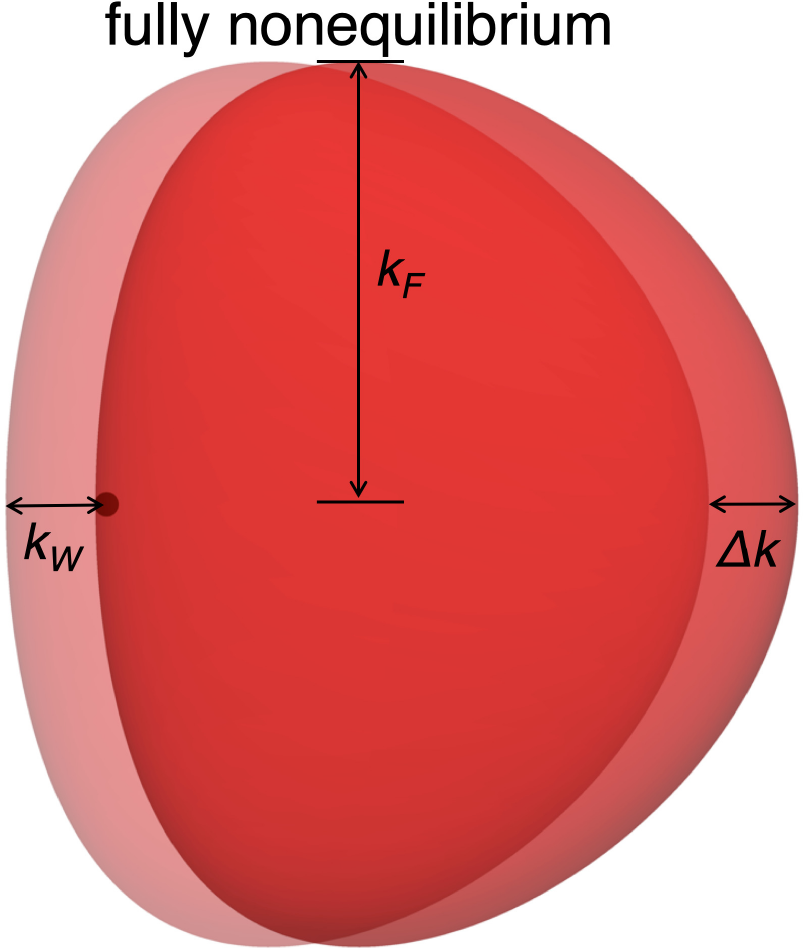}\\[0.5cm]
\includegraphics[width=0.45\textwidth]{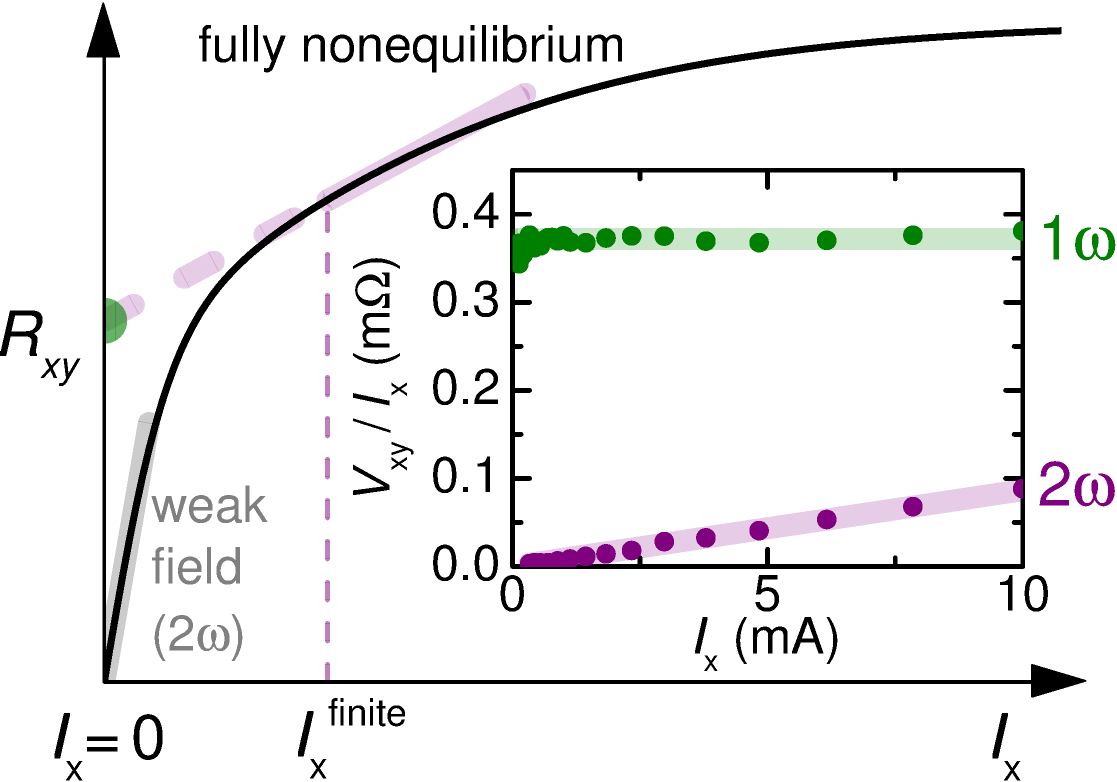}\\[0.2cm]

\end{center}
\vspace{-21.8cm}

\hspace{-8cm}{\bf\large A}

\vspace{4.4cm}

\hspace{-8cm}{\bf\large B}

\vspace{4.6cm}

\hspace{-8cm}{\bf\large C}

\vspace{4cm}

\hspace{-8cm}{\bf\large D}

\vspace{5.cm}

\caption{\label{Fig4} {\bf Theoretical description of Weyl-Kondo physics in
Ce$_3$Bi$_4$Pd$_3$.} ({\bf A}) {\em Ab initio} bandstructure of
Ce$_3$Bi$_4$Pd$_3$, with $4f$ electrons in the core. In the $k_{\rm x}$-$k_{\rm
z}$ plane, four different Weyl nodes (1-4) are identified. 1 and 4 are most
strongly tilted (Supplementary Information Sect.\,\ref{SIDFT}). ({\bf B})
Dispersion across a pair of Weyl ($\rm{W}^+$) and anti-Weyl ($\rm{W}^-$) nodes
for a Weyl-Kondo model with tilted Weyl cones (Supplementary Information
Sect.\,\ref{SImodel}). Energy is expressed in units of the conduction electron
bandwidth $t$. The Kondo interaction pushes the Weyl nodes, that are present }
\end{figure}

We also measured the Hall effect in finite applied magnetic fields. We observe
that the field-dependent Hall resistivity isotherms, $\rho_{xy}(B)$, are
fundamentally different below (Fig.\,\ref{Fig3}A) and above $T^{\star}$
(Fig.\,\ref{Fig3}B). Whereas above $T^{\star}$, $\rho_{xy}$ shows simple
linear-in-field behavior consistent with a single hole-like band, strong
nonlinearities appear below $T^{\star}$. Most \SBP{importantly,} a large even-in-field
component $\rho_{xy}^{\rm{even}} = [\rho_{xy}(B)+\rho_{xy}(-B)]/2$ is observed
(Fig.\,\ref{Fig3}C) that even overwhelms the usual odd-in-field component
$\rho_{xy}^{\rm{odd}} = [\rho_{xy}(B)-\rho_{xy}(-B)]/2$ (Fig.\,\ref{Fig3}E). The
nonlinear part of the latter, that scales with
$[\rho_{xx}-\rho_{xx}(4\,\text{K})]^2$ and is thus independent of the scattering
time (Fig.\,\ref{Fig3}F), is theoretically expected \cite{Nag10.1} and
experimentally observed \cite{Liu18.1} in TRS broken Weyl semimetals, which is
here realized by the finite magnetic field. The exciting new result, however, is
the even-in-$B$ component, which is the finite-field extension of the
spontaneous Hall effect. Both are incompatible with the standard (magnetic field
or magnetization induced) Hall conductivity mechanism, where the elements
$\sigma_{xy}$ of the fully antisymmetric Hall conductivity tensor may couple
only to a physical quantity $G$ that breaks TRS (i.e., $TG=-G$, where $T$ is the
time reversal operation) \cite{Cas45.1} and thus have to be an odd function of
this quantity [e.g., $\sigma_{xy}(B) = -\sigma_{xy}(-B)$ where $G = B$]. 


The question then is how to understand the Hall response beyond \SBP{a broken TRS} framework?
Recent theoretical studies \cite{Sod15.1} show that in an IS breaking but
TRS preserving material, a Hall current (density)\vspace{-0.4cm}

\begin{equation}
j_y = \sigma_{xy}\mathcal{E}_x = \frac{e^2}{\hbar} \int \frac{d^3k}{(2\pi)^3} f(\bm{k}) \underbrace{\Omega^{\rm{odd}}_z(\bm{k})\mathcal{E}_x}_{v_y}
\label{Eq1}
\end{equation}
\noindent can be generated in a current-carrying state, as nonlinear response to
an applied electric field $\mathcal{E}_x$. This field changes the equilibrium
(Fermi-Dirac) distribution function $f_0(\bm{k})$ into the nonequilibrium distribution function $f(\bm{k})$. In addition, in the presence of the
Berry curvature $\Omega^{\rm{odd}}_z(\bm{k})$, that is odd in $\bm{k}$
[i.e., $\bm{\Omega}^{\rm{odd}}(\bm{k}) = - \bm{\Omega}^{\rm{odd}}(-\bm{k})$] for
systems with broken IS \cite{Xia10.1}, it generates the anomalous
velocity $v_y$. This state breaks TRS at the thermodynamic level, as
$f(\bm{k})$ can be maintained only at the cost of entropy production ($\dot{S} = j_x \mathcal{E}_x$). The Hall conductivity $\sigma_{xy}$ in Eqn.\,\ref{Eq1} is finite only under this condition, which is in contrast to a normal (linear-response) Hall conductivity (Supplementary Information Sect.\,\ref{TRS_breaking}).

Because $\sigma_{xy}$ \SBP{of Eqn.\,\ref{Eq1}} is driven by \SBP{this TRS invariant} Berry curvature and not by an applied
magnetic field, it does not need to be odd in $B$, and a finite $\sigma_{xy}$
does not even require the presence of any $B$ at all (thus the spontaneous Hall
effect). In fact, the only influence the magnetic field has on this topological
Hall effect is to successively reduce its magnitude with increasing field
(Fig.\,\ref{Fig3}D), similar to what happens when heating the material beyond
$T^{\star}$ (Fig.\,\ref{Fig3}C). The observations of a spontaneous Hall effect
(Fig.\,\ref{Fig2}) and an even-in-field Hall conductivity (Fig.\,\ref{Fig3}) in
Ce$_3$Bi$_4$Pd$_3$ are smoking-gun evidences that the physical quantity
underlying the phenomenon is not a magnetic order parameter (coupled linearly to
$B$), as otherwise $\sigma_{xy}$ would necessarily be completely odd in $B$.
That the spontaneous Hall current is indeed carried by $f(\bm{k})$ is further
supported by the linear relationship between $\sigma_{xy}$ and $\sigma_{xx}$ in
the Kondo coherent regime (Fig.\,\ref{Fig2}B), consistent with a linear
dependence on scattering time ($\sigma_{xy} \sim \tau$) and thus the
nonequilibrium nature of the effect. As recently emphasized \cite{Du19.1}, this
dependence sharply discriminates this effect from disorder-induced
contributions.

Because according to Eqn.\,\ref{Eq1} the topological Hall current is \SBP{determined} by
$f(\bm{k})$, $\sigma_{xy}$ will depend on $\mathcal{E}_x$, and thus the Hall
response $j_y = \sigma_{xy}(\mathcal{E}_x)\cdot \mathcal{E}_x$ is expected to be
nonlinear in $\mathcal{E}_x$. By Taylor expanding $f(\bm{k})$ around $f_0(\bm{k})$, a second harmonic
response was derived \cite{Sod15.1}, which we have set out to probe by
investigating both the dependence of the Hall response on the electric field (or
current) strength and by analyzing the different components of the Hall response
under AC \SBP{and DC} current drive\SBP{s (Supplementary Information Sect.\,\ref{nonlinearRH}). As we will show in what follows, we do indeed observe this second harmonic response; in addition, however, we also detect terms that go beyond the prediction.} The experiments \SBP{as function of DC current drive} reveal that the spontaneous Hall voltage $V_{xy}^{\rm DC}$
can be
decomposed into a linear- and a quadratic-in-$I^{\rm DC}$ contribution \SBP{(Fig.\,\ref{dc}A\,right).} This is
in contrast to the longitudinal voltage $V_{xx}^{\rm DC}$ that is linear in
current, representing an Ohmic electrical resistivity (Fig.\,\ref{dc}A\,left).
In our AC experiments, in response to an excitation at frequency $\omega$ we
detect voltage contributions at $1\omega$, $2\omega$, and $0\omega$. Whereas
$V_{xy}^{1\omega}$ is linear in $I^{1\omega}$ (Fig.\,\ref{ac1omega}B), both
$V_{xy}^{2\omega}$ \SBP{and the associated current rectified counterpart} $V_{xy}^{0\omega}$ \SBP{(Eqn.\,\ref{Eqn_jHall})} are quadratic in $I^{1\omega}$ (Fig.\,\ref{Fig2}D,\,E). These three responses as well as the DC response
described above appear simultaneously, as Kondo coherence develops with
decreasing temperature (Fig.\,\ref{Fig2}F), and must thus have a common origin.

\noindent{\bf DISCUSSION}\\
\SBP{We start by discussing the terms $V_{xy}^{2\omega}$, $V_{xy}^{0\omega}$,
and $V_{xy}^{\rm DC} \sim I^2$. They can be understood within the perturbative
treatment \cite{Sod15.1} of Eqn.\,\ref{Eq1}, where the spontaneous nonlinear
Hall conductivity $\sigma_{xy}$ is determined by the Berry curvature dipole $D_{xz}$ as
\vspace{-0.4cm}

\begin{equation}
\sigma_{xy} = \frac{e^3\tau}{\hbar^2}\cdot D_{xz}\cdot\mathcal{E}_x \quad \mbox{with} \quad D_{xz} = \int \frac{d^3 k}{(2\pi)^3} f_0(\bm{k}) \frac{\partial\Omega^{\rm{odd}}_z}{\partial k_x} \; .
\label{Eq2}
\end{equation}
\noindent For several noninteracting (and TRS preserving) Weyl semimetals,
$D_{xz}$ was computed  by electronic bandstructure calculations \cite{Zha18.1}.}
The tangent of the Hall angle, defined as $\tan\Theta_{\rm H} \equiv
\sigma_{xy}/\sigma_{xx}$, where $\sigma_{xx}$ is the normal ($1\omega$)
longitudinal conductivity, was found to be at maximum \SBP{(if the chemical
potential is placed at the Weyl nodes)} of the order $10^{-4}$ for a scattering
time $\tau = 10$\,ps and an electric field of $\mathcal{E}_x = 10^2$\,V/m,
corresponding to $\tan\Theta_{\rm H}/\mathcal{E}_x \le 10^{-6}$\,m/V
\cite{Zha18.1}. \SBP{An experimental confirmation of these predictions remaines
elusive to date.} For Ce$_3$Bi$_4$Pd$_3$, we have measured $\tan\Theta_{\rm
H}/\mathcal{E}_x$ values as large as $3\times 10^{-3}$\,m/V in the second
harmonic channel (Fig.\,\ref{Fig2}D). \SBP{This giant value} is even more
surprising as it is obtained in a bulk semimetal, without any chemical potential
tuning. Experimentally, such tuning is limited to the case of \SBP{insulating} 
(quasi) 2D materials.

\SBP{Two comments are due. Firstly, in gate-tuned} bilayer \cite{Ma19.1} and
few-layer \cite{Kan19.1} WTe$_2$, \SBP{which both feature a gap at the Fermi
level,} $2\omega$ Hall voltages have recently been \SBP{reported; they were
attributed to a large---but not divergent---Berry curvature.} We estimate the
maximum values reached there to be $\tan\Theta_{\rm H}\approx 5\times 10^{-3}$
and $10^{-4}$, and $\tan\Theta_{\rm H}/\mathcal{E}_x \approx 3 \times
10^{-6}$\,m/V and $10^{-8}$\,m/V, respectively, thus again at least three orders
of magnitude smaller than what we observe \SBP{in Ce$_3$Bi$_4$Pd$_3$. Secondly,
we point out that the above-discussed spontaneous nonlinear Hall effect is not
to be confused with the anomalous Hall effect in TRS breaking Weyl semimetals
\cite{Liu18.1} or the planar Hall effect in noncentrosymmetric Weyl semimetals
\cite{Nan17.1}. In the former, the Berry curvature is related to a
magnetization. In the latter, the amplitude of the effect is determined by that
of the chiral anomaly \cite{Nan17.1}, which is strongly suppressed in Weyl-Kondo
semimetals.}

\SBP{Next we quantify the terms $V_{xy}^{1\omega}$  and $V_{xy}^{\rm DC} \sim I$
that} have not been considered in the perturbative approach of
ref\,\cite{Sod15.1}, though odd-in-current contributions can generally
appear (Supplementary Information Sect.\,\ref{OddInCurrent}). Intriguingly, these linear contributions show
$\tan\Theta_{\rm H}$ values up to to 0.5 (taken as the slope
$\partial\sigma_{xy}/\partial\sigma_{xx}$ in Fig.\,\ref{Fig2}B and
Fig.\,\ref{reproduce}B), \SBP{and thus are} even larger than the $2\omega$
effect quantified above. \SBP{Before describing how these new terms as well as
the above-discussed spontaneous nonlinear Hall terms may naturally arise in a
Weyl-Kondo semimetal picture, we show that other effects can be safely
discarded.}

\SBP{Skew scattering, side jump, and multiband effects are investigated via the normal (antisymmetrized) Hall effect and magnetoresistance characteristics, and shown to play negligible roles by quantitative analyses (Supplementary Information Sects.\ref{skew}, \ref{multi} and Figs.\,\ref{SMfig1}, \ref{SMfig3}). Spurious Hall contributions due to crystal anisotropies, which may arise in systems with lower than cubic symmetry, can also be ruled out (Supplementary Information Sect.\ref{trivialRH}). Finally, the reproducibility of the spontanous Hall effect over various samples (Fig.\,\ref{reproduce}) confirms the intrinsic nature of this phenomenon.}

\SBP{In a Weyl-Kondo semimetal, the Weyl nodes---where the Berry curvature is singular---are} essentially
pinned to the Fermi level \cite{Dzs17.1,Lai18.1}. The application of even
small electric fields has a nonperturbative effect, in particular in the case of tilted Weyl cones (Supplementary
Information Sect.\,\ref{SINLH} for an expanded discussion). A simple Taylor expansion of
$f(\bm{k})$ around $f_0(\bm{k})$, as done in ref\,\cite{Sod15.1}, will
therefore fail to describe the situation. Quantitative predictions from fully
nonequilibrium transport calculations based on an {\em ab initio} electronic
bandstructure in the limit of strong Coulomb interaction and strong spin-orbit
coupling are elusive to date. Thus, instead, we here present a conceptual
understanding. While tilted Weyl cones are present already in the
noninteracting bandstructure of Ce$_3$Bi$_4$Pd$_3$ (Fig.\,\ref{Fig4}A and
Supplementary Information Sect.\,\ref{SIDFT}), it is the Kondo interaction that 
\qs{drives emergent and highly renormalized Weyl nodes in the} immediate vicinity of the Fermi level (Fig.\,\ref{Fig4}B), as indicated by
calculations for a periodic Anderson \qs{model \cite{Lai18.1}} with tilted Weyl cones in the bare
conduction electron band \qs{(Supplementary
Information Sect.\,\ref{SImodel}, Fig.\,\ref{Fig:Dispersion_TWKSM})} and evidenced by thermodynamic measurements \cite{Dzs17.1}. In the resulting tilted Weyl-Kondo semimetal, each Weyl node will be
asymmetrically surrounded by a small Fermi pocket. The presence of such small pockets in Ce$_3$Bi$_4$Pd$_3$ is supported by the small carrier concentration ($8 \times 10^{19}$\,cm$^{-3}$, or 0.002 charge carriers per atom) obtained from the normal Hall coefficient (Fig.\,\ref{SMfig1}A). Due \SBP{to the pinning of the Fermi energy to the Weyl nodes in a tilted Weyl-Kondo semimetal,}
the smallest distance
between node and Fermi surface ($k_{\rm W}$, see Fig.\,\ref{Fig4}C) can become extremely small. The
application of even tiny electric fields will then induce shifts $\Delta\bm{k}$
in the distribution function $f(\bm{k})$ that are sizeable compared to $k_{\rm
W}$ (and possibly even compared to the Fermi wavevector $k_{\rm F}$), thus
driving the system to a fully nonequilibrium regime (Fig.\,\ref{Fig4}D). In this
setting, a nonperturbative approach is needed, and will allow for the appearance
of terms beyond the second harmonic one predicted by Sodemann and Fu
\cite{Sod15.1}, most notably the experimentally observed first harmonic one
(Supplementary Information Sect.\,\ref{nonlinearRH}). As the applied $\mathcal{E}$ field can then no
longer be considered as a probing field, it introduces a directionality on top
of the crystal's space group symmetry and the selection rules, that hold in the
perturbative regime, will be violated.


In conclusion, our Hall effect measurements have unambiguously identified a
giant Berry curvature contribution in a time-reversal invariant material, the
noncentrosymmetric heavy fermion semimetal Ce$_3$Bi$_4$Pd$_3$. The Hall angle
per applied electric field, a figure-of-merit of the effect, is enhanced by
orders of magnitude over values expected for weakly interacting systems, which
we attribute to \SBP{the effect of tilted} \qs{and highly renormalized} \SBP{Weyl nodes}
\qs{that emerge} very close to the Fermi surface \qs{out of the Kondo effect.} The experiments established here should allow for a ready identification of 
other strongly correlated nonmagnetic Weyl semimetals, be it in heavy fermion
compounds or in other materials classes, thereby enabling much needed systematic
studies of the interplay of strong correlations and topology. Our findings provide a window into the landscape of ``extreme topological matter''---where strong correlations lead to extreme topological responses---that awaits systematic exploration. Finally, the
discovered effect being present in a 3D material, in the absence of any magnetic
fields and under only tiny driving electric fields, holds great promise for the
development of robust topological quantum devices.

\newpage



%

\vspace{0.5cm}

\noindent{\bf ACKNOWLEDGEMENTS}\\
The authors thank J.-C.\ Orain for technical assistance
during the $\mu$SR experiments, C. Wilhelmer for contributions to transport
experiments, and J.\ Mesot and D.\ A.\ Zocco for fruitful discussion.
The team in Vienna acknowledges financial support from the Austrian
Science Fund (FWF grants No.\ P29279-N27, P29296-N27, and DK W1243) and the
European Union's Horizon 2020 Research and Innovation Programme, under Grant
Agreement no EMP-824109. T.S.\ acknowledges support from the Swiss National
Science Foundation (SNF Grant No.\ 200021-169455). Work at Rice was in part
supported by the NSF (DMR-1920740) and the Robert A.\ Welch Foundation (C-1411),
and by a Ulam Scholarship from the Center for Nonlinear Studies at Los Alamos
National Laboratory.\\

\noindent{\bf AUTHOR CONTRIBUTIONS}\\
S.P.\ initiated and lead the
study. S.D., X.Y., and A.P.\ synthesized and characterized the material. S.D.\
and M.T.\ performed and analyzed the magnetotransport measurements. G.E.\
contributed to the data analysis. T.S.\ and S.D.\ performed the $\mu$SR
investigation. P.B.\ and O.R.\ performed the {\em ab initio} study, S.E.G.,
H.H.L., and Q.S.\ the study of the Weyl-Kondo semimetal model. S.D.\ and S.P.\
prepared the manuscript, with input from all authors. All authors contributed to
the discussion.\\

\noindent{\bf COMPETING INTERESTS}\\
The authors declare that they have no
competing interests.\\

\noindent{\bf DATA AND MATERIALS AVAILABILITY}\\
All data needed to
evaluate the conclusions in the paper are present in the paper and/or the
Supplementary Notes and Figures. Additional data related to this paper may be requested
from the authors.

\newpage



\newpage


\newpage


\clearpage

\newpage


\clearpage
\newpage

\begin{center}
cont. FIG.\ 4: ...  in the bare conduction electron band far away from the Fermi
energy, to the immediate vicinity of the Fermi level (here at $E/t = 0$,
slightly above the Weyl nodes). ({\bf C})~Sketch of a Fermi pocket around the
Weyl node $\rm{W}^+$ in ({\bf B}) (dot) in zero electric field (light red) and
its nonequilibrium counterpart with driving electric field (red), in the
weak-field regime (left) and the fully nonequilibrium regime (right). ({\bf D})
Sketch of the driving current-induced Hall resistance $R_{xy} = V_{xy}/I_x$,
displaying the weak-field and fully nonequilibrium regimes. The inset shows
experimental results for Ce$_3$Bi$_4$Pd$_3$, demonstrating the simultaneous
presence of a spontaneous Hall signal in both the $2\omega$ ($V_{xy} \sim
I_x^2$) and $1\omega$ ($V_{xy} \sim I_x$) channel. The sum of both contributions
corresponds to a linear-in-$I_x$ Hall resistance with a finite offset (violet
line with green dot in main panel), which is a characteristic of the fully
nonequilibrium regime.
\end{center}


\label{SIHall} 
\label{misaligne} 
\label{SIeq1} 
\label{SMfig_ContactSketch} 
\label{eq_mis} 
\label{SMfig2} 
\label{Fig_Rxx_corr} 
\label{skew} 
\label{Eq:SM1} 
\label{SMfig1} 
\label{Eq:SM2} 
\label{multi} 
\label{SMfig3} 
\label{nonlinearRH} 
\label{Hallcurrent} 
\label{Eqn_jHall} 
\label{eq_defs} 
\label{dc} 
\label{ac1omega} 
\label{SMtable} 
\label{reproduce} 
\label{TRS_breaking} 
\label{OddInCurrent} 
\label{Eqn_SHE_sigma} 
\label{Eqn_SHE_Sod} 
\label{Eqn_FermiShift1} 
\label{Eqn_SHE_Sod2} 
\label{trivialRH} 
\label{Eq:ja1} 
\label{SI_RxxJump} 
\label{Eqn_Rxx_drop} 
\label{Rxx_jump} 

\label{SImuSRCp} 
\label{SImuSR} 
\label{eq:KT_and_electr} 
\label{fig:ZF_LF_muSR} 
\label{SImagnetization} 
\label{Fig:SI_hysteresis} 
\label{SICp} 
\label{fig:Cp_S1} 

\label{SIDFT} 
\label{SMtable1} 

\label{SImodel} 
\label{WKSM} 
\label{PAM1} 
\label{eq:hc_sigma} 
\label{eq:hc_sigma} 
\label{eq:hybridization}
\label{eq:hd} 
\label{eq:d1} 
\label{eq:d2} 
\label{eq:Dx} 
\label{Fig:Dispersion_TWKSM} 
\label{WKSMlow} 
\label{j2omega2} 
\label{Eq:BCD_FS2} 
\label{Eq:Dxy_prod_form} 
\label{Fig:tilted_nodes} 
\label{SINLH} 
\label{Delta-vy} 
\label{Delta-vz} 
\label{totalf} 
\label{BZ} 
\label{eq:1omega} 
\label{eq:jy} 

\newpage

\noindent{\bf Methods}

\noindent{\bf A. Synthesis} 

\noindent Single crystals of Ce$_3$Bi$_4$Pd$_3$ and the nonmagnetic reference
compound La$_3$Bi$_4$Pd$_3$ where synthesized using a Bi-flux method
\cite{Dzs17.1}. Their stoichiometry, phase purity, and crystal structure were
verified using powder X-ray diffraction, scanning electron microscopy, energy
dispersive X-ray spectroscopy, and Laue diffraction. \SBP{Because Ce$_3$Bi$_4$Pd$_3$ is a stoichiometric compound in which all three elements have unique crystallographic sites, disorder is expected to be weak.}

\noindent{\bf B. Measurement setups} 

\noindent Magnetotransport measurements were performed using various devices:
two Quantum Design Physical Property Measurement Systems, in part with $^3$He or
vertical rotator option, and an Oxford $^4$He flow cryostat using a Stanford
Research SR830 lock-in amplifier. In the former, we used a pseudo-AC technique (p-AC), in the latter a standard AC technique with lock-in detection. Electrical contacts for these measurements
where made by spot welding 12\,$\mu$m diameter gold wires to the samples in a
5- or 6-wire configuration, depending on the crystal size. Oriented single
crystals were studied with the driving electrical current along different
crystallographic directions (approximately along $[103]$, $[111]$, and
$[100]$).

The $\mu$SR measurements were performed at the Dolly spectrometer of the Swiss
Muon Source at Paul Scherrer Institut, Villigen. The single crystals were
arranged to form a mosaic with about 1\,cm diameter and a thickness of about
0.5\,mm, glued on top of a thin copper foil solidly clamped to a copper sample
holder, thus optimally using the muon beam cross section, minimizing the
background from the sample holder, and guaranteeing good thermal contact.
Combined with active vetoing, this setup resulted in very low spurious
background signals. A cold-finger Oxford Heliox $^3$He system combined with a
$^4$He Oxford Variox cryostat was used to reach temperatures down to 250\,mK. By
employing active compensation coils, true zero-field (ZF) conditions could be
achieved during the ZF-$\mu$SR experiments.

\noindent{\bf C. {\em Ab initio} calculations}

\noindent We performed nonspinpolarized bandstructure calculations for
Ce$_3$Bi$_4$Pd$_4$ based on density functional theory, treating the Ce $4f$
electrons in the open-core approximation and taking spin-orbit interaction into
account. Weyl nodes in the $k_x$-$k_z$ plane of the Brillouin zone were
identified via their Berry curvature (Supplementary Information Sect.\,\ref{SIDFT}).

\noindent{\bf D. Model calculations}

\noindent We extended the model for a Weyl-Kondo semimetal \cite{Lai18.1} to
include beyond nearest-neighbor hopping terms, and solved the self-consistent
saddle-point equations for the strong interaction limit of the periodic Anderson
model. We find a Weyl-Kondo solution with tilted Weyl cones. With the Kondo
interaction placing the Fermi energy very close to the Weyl nodes, the Fermi
surface comprises Fermi pockets that are asymmetrically distributed near the
Weyl and anti-Weyl nodes. The Berry curvature, which diverges exactly at any
Weyl or anti-Weyl node, is thus very large on the Fermi surface (see
Supplementary Information Sect.\,\ref{SImodel} for further details).

\newpage



\addtocounter{figure}{-4} 
\addtocounter{equation}{-2} 
\makeatletter
\renewcommand{\thefigure}{S\@arabic\c@figure}
\renewcommand{\theequation}{S\@arabic\c@equation}
\renewcommand{\thetable}{S\@arabic\c@table}

\renewcommand{\theHtable}{Supplement.\thetable}
\renewcommand{\theHfigure}{Supplement.\thefigure}

\noindent{\large\bf Supplementary Information for}
\vspace{1cm}
\setcounter{page}{1}

\noindent{\bf Giant spontaneous Hall effect in a nonmagnetic Weyl-Kondo semimetal}
\vspace{0.5cm}

\noindent S. Dzsaber, X.\ Yan, M.\ Taupin, G.\ Eguchi, A.\ Prokofiev, T.\
Shiroka, P. Blaha, O.\ Rubel, S.\ E.\ Grefe, H.-H.\ Lai, Q.\ Si, and S.~Paschen
\vspace{0.5cm}

\noindent Corresponding Author: Silke Paschen
\vspace{0.1cm}

\noindent Email: paschen@ifp.tuwien.ac.at
\vspace{1cm}

\noindent {\bf This PDF file includes:}
\vspace{0.5cm}


Supplementary Text Sects.\,I to IV
\vspace{0.1cm}

Figures S1 to S15
\vspace{0.1cm}

Tables S1 and S2
\vspace{0.1cm}

Equations S1 to S36


\newpage



\noindent{\bf\large Supplementary Text}

\section{Analysis of Hall effect and resistivity data}\label{SIHall}
\subsection{Magnetotransport measurement geometry}\label{misaligne}
\noindent We define the electrical current direction as $x$. The electrical
resistivity along this direction is thus denoted by $\rho_{xx}$. A magnetic
field applied along $x$ leads to longitudinal magnetoresistance, a field
perpendicular to $x$ to transverse magnetoresistance. To detect a normal Hall
response, a magnetic field is applied along $z$, and an electric field along $y$
is measured via voltage contacts perpendicular to $x$ and $z$. 

However, in practice, by contacting the (small) samples with the spot welding
technique (Methods), this geometry cannot be perfectly achieved, leading to
misalignment contributions for both the resistance and the Hall effect
measurements (Fig.\,\ref{SMfig_ContactSketch}). We start by discussing the
latter. We denote the resistance measured across imperfectly aligned Hall
contacts as $R^{\rm{meas}}_{xy}$, the genuine Hall resistance as $R_{xy}$. The
established technique to cancel out misalignment contributions in the normal
Hall effect is to measure $R^{\rm{meas}}_{xy}$ for positive and negative
magnetic fields and antisymmetrize the signal as \vspace{-0.4cm}

\begin{equation}
R_{xy} = \frac{R^{\rm{meas}}_{xy}(+B_z)-R^{\rm{meas}}_{xy}(-B_z)}{2} \;\; .
\label{SIeq1}
\end{equation}

\begin{figure}[b!]
\includegraphics[width=0.55\textwidth]{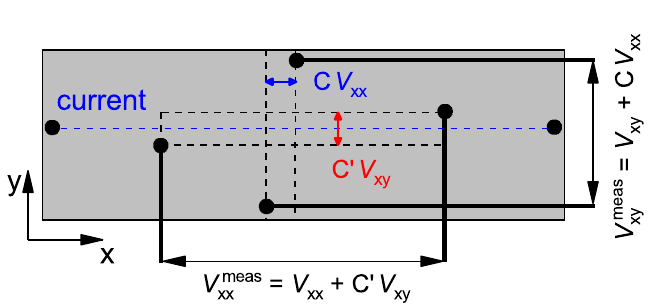}

\caption{\label{SMfig_ContactSketch} {\bf Misalignment contributions due to imperfect contact geometry.} Because of unavoidable contact misalignment, the measured voltages across the (longitudinal) resistivity and Hall contacts  ($V_{xx}^{\rm meas}$ and $V_{xy}^{\rm meas}$) contain, in addition to the intrinsic contributions $V_{xx}$ and $V_{xy}$, the misalignment contributions $C' V_{xy}$ and $C V_{xx}$, respectively.}
\end{figure}

For the spontaneous and even-in-field Hall response, however, a different
approach is needed. At room temperature and in zero applied magnetic field,
where no Hall response exists, we determine the ratio of resistances
measured across the (longitudinal) resistivity and Hall contacts
as\vspace{-0.4cm}

\begin{equation}
C  = \frac{R^{\rm{meas}}_{xy}}{R_{xx}^{\rm{meas}}} = \frac{C R_{xx}}{R_{xx}} = 0.553 \;\; .
\end{equation}
\noindent As the misalignment factor $C$ is a purely geometrical quantity and is
thus temperature and field independent, the genuine Hall response at arbitrary
fields and temperatures is\vspace{-0.4cm}

\begin{equation}
R_{xy}(T,\,B)= R^{\rm{meas}}_{xy}(T,\,B) - C R_{xx}(T,\,B) \;\; .
\label{eq_mis}
\end{equation}
\noindent In fact, in the correction term we use
$R_{xx}^{\rm{meas}}(T,\,B)$ instead of $R_{xx}(T,\,B)$, thus correcting the
effect only to first order. As higher order terms involving the product $CC'$
are very small (for $C'$ see Fig.\,\ref{Fig_Rxx_corr}), this is deemed
sufficiently accurate. $R_{xy}(T,\,B=0)$ is zero from room temperature down to
3\,K (Fig.\,\ref{SMfig2}), thus ruling out that the additional signal measured
across the Hall contacts below 3\,K is due to Hall contact misalignment.

\begin{figure}[b!]
\includegraphics[width=0.45\textwidth]{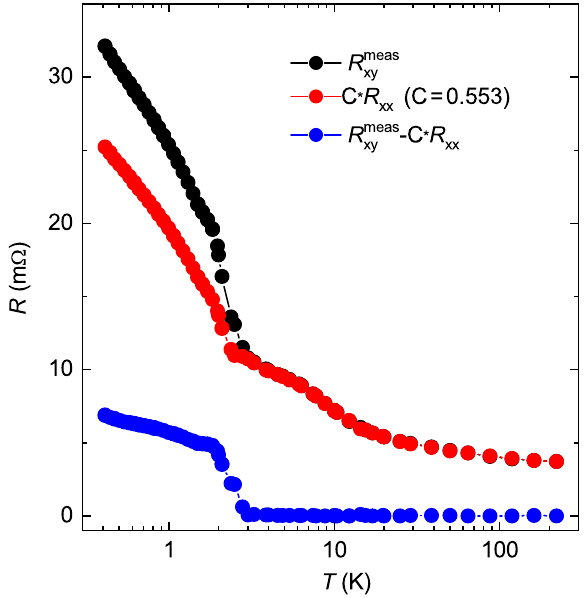}

\caption{\label{SMfig2} {\bf Determination of misalignment contribution in DC Hall measurements
on Ce$_3$Bi$_4$Pd$_3$.} Temperature-dependent electrical resistance measured
across the Hall contacts $R^{\rm{meas}}_{xy}$ (black), longitudinal resistance
$R_{xx}$ scaled to $R^{\rm{meas}}_{xy}$ at room temperature (red), and the
difference of the two, $R_{xy}=R^{\rm{meas}}_{xy}-C\cdot R_{xx}$ with $C =
0.553$ (blue), all in zero applied magnetic field. A deviation appears only
below $T = 3$\,K, where the intrinsic spontaneous Hall response sets in.
The same procedure is adopted for the $1\omega$ signal of the AC Hall
measurements.}
\end{figure}

This Hall contact misalignment correction is adopted for both the DC
transport measurements and the $1\omega$ signal in the AC measurements. As there
is no longitudinal signal in the $0\omega$ and $2\omega$ channels, no correction
is done for these.

We now turn to the electrical resistivity. Here, the transverse direction misalignment of the contacts leads to the appearance of an additional resistance $C' R_{xy}$, that adds to the intrinsic longitudinal resistance $R_{xx}$, so that the total measured resistance is $R^{\rm meas}_{xx} = R_{xx} + C'R_{xy}$ (with $R_{xy}$ obtained via Eqn.\,\ref{eq_mis}). We determine the misalignment factor $C'$ as follows: At a temperature above the onset of the spontaneous Hall effect we measure, on the resistivity contacts, the normal Hall resistance $[R_{xx}^{\rm meas}(+B)-R_{xx}^{\rm meas}(-B)]/2$ as function of magnetic field (Fig.\,\ref{Fig_Rxx_corr}A, black data points).
\begin{figure}[b!]
\centering
\includegraphics*[width=1\textwidth]{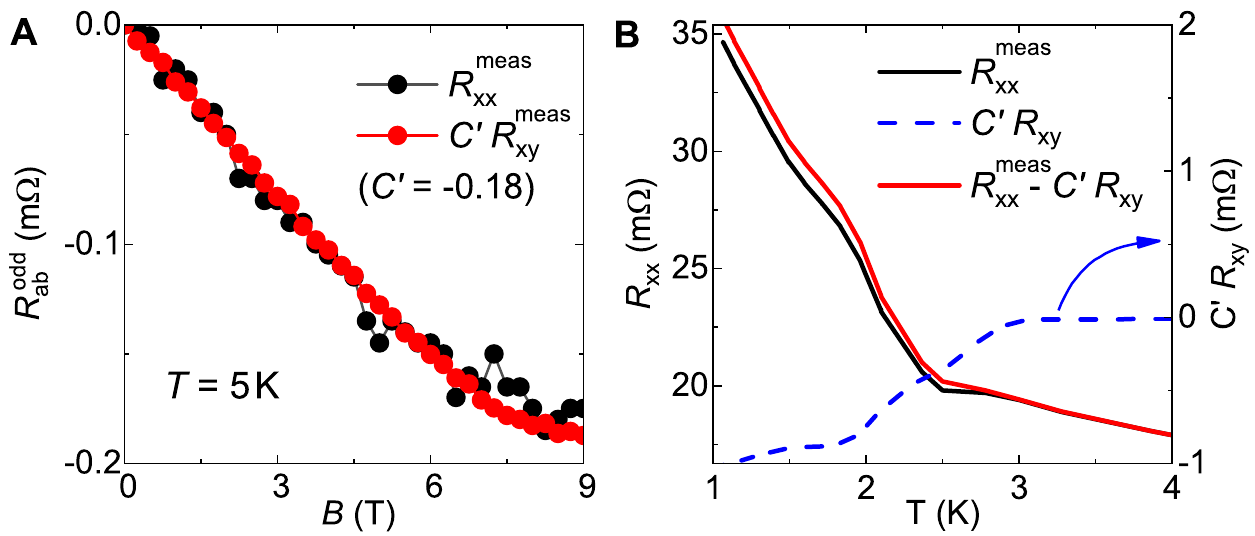}

\caption{\label{Fig_Rxx_corr} {\bf Misalignment correction of the electrical resistance}. ({\bf A}) Measured odd-in-$B$ resistance $R_{ab}^{\rm{odd}}=[R_{ab}^{\rm meas}(+B)-R_{ab}^{\rm meas}(-B)]/2$ on the electrical resistance contacts ($ab = xx$, black), together with that measured across the Hall contacts and scaled by $C'$ ($ab = xy$, red). Data are plotted for sample S1 (see MS), and obtained at $T = 5\,$K, above the onset of the spontaneous Hall effect. $C'$ is chosen to achieve best overlap of the curves. ({\bf B}) With $C'$ obtained from ({\bf A}), the measured temperature dependent resistance $R^{\rm meas}_{xx}$ (black) is corrected for the misalignment contribution $C'R_{xy}$ (blue dashed line), to obtain the intrinsic curve (red).}
\end{figure}
At the same temperature, we also measure the Hall resistance on the Hall contacts, and scale it (by the factor $C'$) such that it collapses onto the former curve (Fig.\,\ref{Fig_Rxx_corr}A, red data points). With $C'$ determined in this way, the temperature dependences of $C'R_{xy}$ and thus of $R_{xx}$ can be determined in zero magnetic field, across the onset of spontaneous Hall effect (Fig.\,\ref{Fig_Rxx_corr}B). One sees that, with this correction (red curve), the feature in $R_{xx}(T)$ seen at the onset of the spontaneous Hall effect is weakened, but still visible (see Fig.\,\ref{Rxx_jump}A). The same is true for the other samples studied in this work (Fig.\,\ref{Rxx_jump}B,\,C). The origin of this additional effect, which is an intrinsic imprint of the Hall conductivity via the giant Hall angle of Ce$_3$Bi$_4$Pd$_3$, is discussed in Sect.\,\ref{SI_RxxJump}.

Finally, if we do this same correction for the 
$2\omega$ longitudinal voltage signal $V_{xx}^{2\omega}$, we find that the signal essentially cancels (Fig.\,\ref{Fig2}D). This is because there is no $2\omega$ signal in the (longitudinal) resistance.

\subsection{\SBP{Skew-scattering and side-jump contributions to the Hall effect}}\label{skew}

\noindent In heavy fermion metals, the Kondo interaction may lead to
non-negligible skew scattering \cite{Fer87.1,Hun04.1,Pas04.1,Nag10.1,Cus12.1}.
The (linear-response) Hall coefficient in the Kondo incoherent regime then reads
\vspace{-0.4cm}

\begin{equation}
R_H(T) = R^0_H + C_1 \rho_{\rm{mag}}(T) \chi(T) \;\; ,
\label{Eq:SM1} 
\end{equation}
\noindent where $R^0_{\rm{H}}$ is the normal Hall coefficient due to charge
carriers [in a simple one-band metal with a temperature-independent charge
carrier density $n$, $R^0_{\rm{H}} = 1/(ne)$], $\rho_{\rm{mag}}$ is the magnetic
contribution to the electrical resistivity (defined as $\rho_{\rm{mag}} =
\rho^{\rm{Ce}}_{xx} - \rho^{\rm{La}}_{xx}$, where $\rho^{\rm{Ce}}_{xx}$ and
$\rho^{\rm{La}}_{xx}$ are the resistivity of the Ce-based heavy fermion compound
and its nonmagnetic La-based reference compound, respectively), $\chi$ is the
magnetic susceptibility, and $C_1$ is a temperature-independent constant.

The linear-response normal Hall coefficient $R_{\rm{H}}$ of Ce$_3$Bi$_4$Pd$_3$
is obtained from the total antisymmetrized Hall signal (Eqn.\,\ref{SIeq1}) at
high temperatures, and as $\rho^{\rm{odd}}_{xy}(8\,{\rm T})/(8\,{\rm T})$ at
4\,K and below (see Fig.\,\ref{Fig3}E of the main part). It shows significant
temperature dependence at high temperatures, but tends to saturate at low
temperatures (Fig.\,\ref{SMfig1}A), in agreement with data above 2\,K from ref\,\cite{Kus19.1}, which is typical of a Kondo semimetal. 
Nevertheless, we here explore whether skew scattering might alternatively lead
to this temperature dependence. For this purpose, we plot $R_{\rm{H}}$ vs
$\rho_{\rm{mag}} \chi^* = \rho_{\rm{mag}}\chi/C$, with temperature as an
implicit parameter (Fig.\,\ref{SMfig1}B), where $C =
n_{\rm{Ce}}\mu_{\rm{eff}}^2/(3 k_{\rm{B}})$ is the Curie constant ($n_{\rm{Ce}}$
is the density and $\mu_{\rm{eff}} = 2.54 \mu_{\rm{B}}$ the effective moment of
the Ce$^{3+}$ ions, $k_{\rm{B}}$ is the Boltzmann constant). In the fully
incoherent regime above 10\,K, a linear relationship with the slope $C_1 =
0.38$\,K/T is indeed observed (Fig.\,\ref{SMfig1}B). At first, this seems to
suggest that the Hall response of this material is dominated by skew scattering.
However, as outlined in what follows, the obtained value of $C_1$ is
unphysically large, indicating that $R_{\rm{H}}(T)$ is dominated by the
temperature dependence of the charge carrier concentration and not by skew
scattering.

\begin{figure}[tb!]
\includegraphics[width=0.9\textwidth]{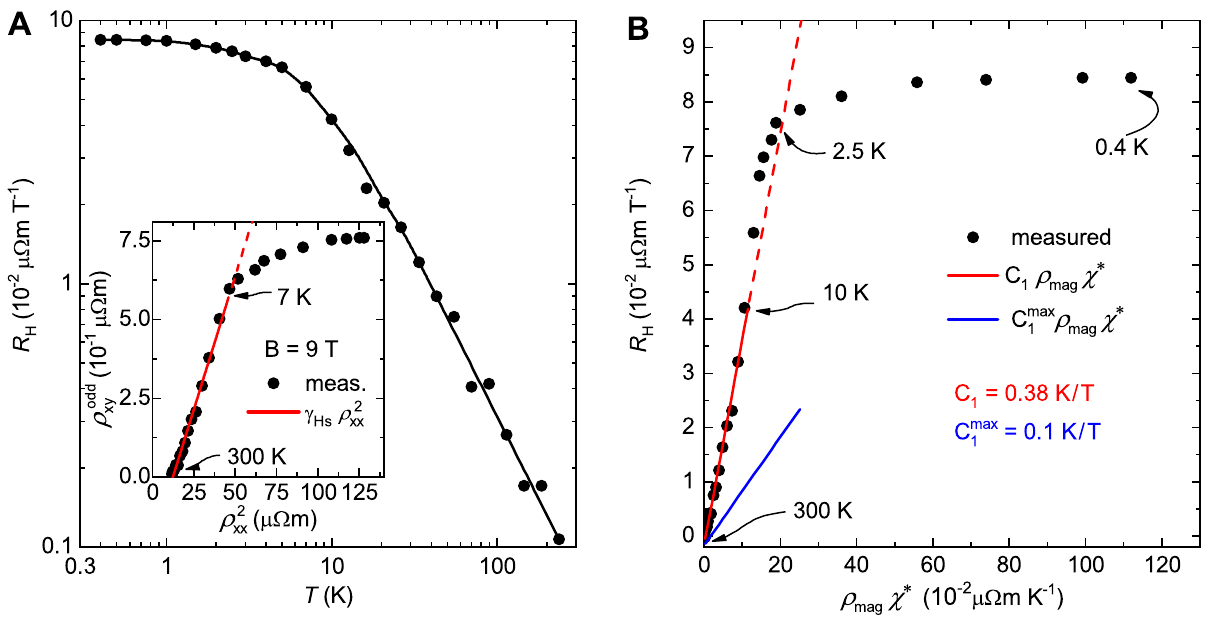}

\caption{\label{SMfig1} {\bf Skew scattering \SBP{and side-jump effect} vs normal Hall effect in Ce$_3$Bi$_4$Pd$_3$.} ({\bf
A}) The linear-response normal Hall coefficient $R_{\rm{H}}$ of
Ce$_3$Bi$_4$Pd$_3$ shows pronounced temperature dependence in the incoherent
regime. \SBP{The inset shows the corresponding $\rho^{\rm odd}_{xy}$ vs $\rho^2_{xx}$ data. The red line indicates a linear fit between 300 and 7\,K with the slope $\gamma_{\rm Hs} = 0.017\,\Omega^{-1}\mu$m$^{-1}$. This value is too large to be attributed to the side-jump effect (see text)}. ({\bf B}) $R_{\rm{H}}$ vs $\rho_{\rm{mag}}\chi^{*}$, with $T$ as implicit
parameter (see text), is linear above 10\,K (full red line is a linear fit to
the data, red dashed line its extrapolation), but the slope $C_1$ is too large
for this behavior to be attributed to skew scattering (slope of blue line
corresponds to the upper boundary).}
\end{figure}
 
For Kondo systems the skew scattering amplitude $C_1$ is given by\cite{Fer87.1}
\vspace{-0.4cm}

\begin{equation}
C_1 = -\frac{5}{7} g_J \frac{\mu_{\rm{B}}}{k_{\rm{B}}} \cos(\delta)\sin(\delta) \;\; \mbox{with} \;\; Z = \frac{2}{\pi} \sum_{-l}^{l} (2l+1) \delta \;\; ,
\label{Eq:SM2} 
\end{equation}
\noindent where $\delta$ is the phase shift of the Fermi wavefunction due to
Kondo scattering, that is related to the valence difference $Z$ between the
impurity and the host metal via Friedel's sum rule, $l$ is the orbital quantum
number, and $J$ is the quantum number of the spin-orbit coupled total moment.
Thus, both $\delta$ and $C_1$ have an upper boundaries. For Ce$^{3+}$ Kondo
scatterers ($Z = 1$, $l = 3$), the maximum value $C_1=0.1$\,K/T (blue line in
Fig.\,\ref{SMfig1}B) is obtained by assuming full screening of the $4f$ moment
($\delta = \pi/14$). Whereas experimentally determined $C_1$ values for
prototypal heavy fermion metals (e.g., $C_1 = 0.082$, 0.075, 0.016, and
0.01\,K/T for CeCu$_6$, CeAl$_3$, CeCoRh$_5$, and CeCoIr$_5$,
respectively\cite{Fer87.1,Hun04.1}) are indeed much smaller than this upper
boundary, our $C_1$ value for Ce$_3$Bi$_4$Pd$_3$ strongly overshoots it. In
fact, such a large $C_1$ value is not only incompatible with the Ce$^{3+}$ case,
but inconsistent with Eqn.\,\ref{Eq:SM2} for any phase shift $\delta$, as $|C_1|
\le 0.21$. We therefore conclude that skew scattering plays at best a minor role
in determining $R_{\rm{H}}(T)$ in the incoherent regime.

In the Kondo coherent regime, skew scattering is known to freeze out with
decreasing temperature, any remnant skew scattering amplitude keeping a positive
sign in Ce-based heavy fermion compounds \cite{Fer87.1}. As the anomalous Hall
effect we observe in Ce$_3$Bi$_4$Pd$_3$ (see Fig.\,\ref{Fig3}E of the main part)
increases in magnitude with decreasing temperature and is negative in sign, we
can safely discard any influence of skew scattering also here.

\SBP{An anomalous Hall effect contribution due to the side-jump effect may occur in solids with broken TRS, and is well-known in ferromagnetic phases \cite{Nag10.1}. As such it cannot account for the spontaneous Hall resistivity we observe (Fig.\,\ref{Fig2}\,A). Here we show that, in addition, this mechanism does not play any appreciable role in the finite-field Hall response of \cepd. The side-jump contribution $\rho^{\rm Hs}_{xy}$ to the Hall resistivity is proportional to $\tau^0$ and thus reads \cite{Ber70.1,Nag10.1}
\vspace{-0.4cm}

\begin{equation}\label{SideJump1}
\rho^{\rm Hs}_{xy} = \gamma_{\rm Hs} \, \rho^2_{xx} \;\; ,
\end{equation}
with
\vspace{-0.4cm}

\begin{equation}
\gamma_{\rm Hs} = \frac{ne^2\Delta y}{\hbar k_{\rm F}} \;\; .
\end{equation}

\noindent Here $\Delta y \approx 10^{-11}$\,m is the amplitude of the side jump in the Hall direction after an impurity scattering event \cite{Nag10.1, Ber70.1}, $n$ the carrier density, $e$ the electron charge, and $k_{\rm F}$ the Fermi wave vector. For \cepd, the carrier density at 300\,mK is $n = 8\cdot 10^{25}$\,m$^{-3}$, and we estimate $k_{\rm F}$ (in a simple parabolic, free-electron single-band model)
as $k_{\rm F} = 1.3 \cdot 10^9$\,m$^{-1}$, thus yielding the (hypothetical) value $\gamma_{\rm Hs} = 1.5\cdot 10^{-4}\,\Omega^{-1}\mu{\rm m}^{-1}$. We do observe the dependence of Eqn.\,\ref{SideJump1} in the Kondo incoherent regime above 7\,K (Fig.\,\ref{SMfig1} inset), but with a slope $\gamma_{\rm Hs} = 1.7 \cdot 10^{-2}\,\Omega^{-1}\mu{\rm m}^{-1}$ that is two orders of magnitude larger than the expectation, making it very unlikely that this dependence is due to the side-jump effect. In the Kondo coherent regime below 7\,K, $\rho^{\rm odd}_{xy}$ saturates (Fig.\,\ref{SMfig1} inset) and, thus, side-jump effects are unimportant also there.

Ignoring the above reasoning that side jumps cannot generate a spontaneous Hall
contribution, we nevertheless estimate the (non-spontaneous) Hall angle tangent due to the
side-jump effect. With Eqn.\,\ref{SideJump1} we obtain $\tan{\Theta_{H}} =
\rho^{\rm Hs}_{xy}/\rho_{xx} = \gamma_{\rm Hs}\rho_{xx}$. Using the above
estimate of $\gamma_{\rm Hs}$ and the electral resistivity $\rho_{xx} \approx 20
\,\Omega\mu{\rm m}$ (Fig.\,\ref{Fig1}\,A) yields $\tan{\Theta_{H}}
\approx 3 \cdot 10^{-3}$, which is three orders of magnitude smaller than the
giant measured value of $\tan{\Theta_{H}} \approx 0.5$, confirming that the side-jump effect does not play any appreciable role in \cepd. 

Finally, setting the above quantative considerations aside, there is a fundamental reason why extrinsic contributions should be ruled out to generate the spontaneous Hall response we observe: The measured $\rho_{xy}(B=0)$ is zero above the Kondo coherence temperature $T^\star$ but impurity scattering should be present at all temperatures.}
\vspace{0.5cm}

\subsection{Hall effect from multiple bands}\label{multi}

\begin{figure}[!b]
\begin{center}
\includegraphics[width=0.8\textwidth]{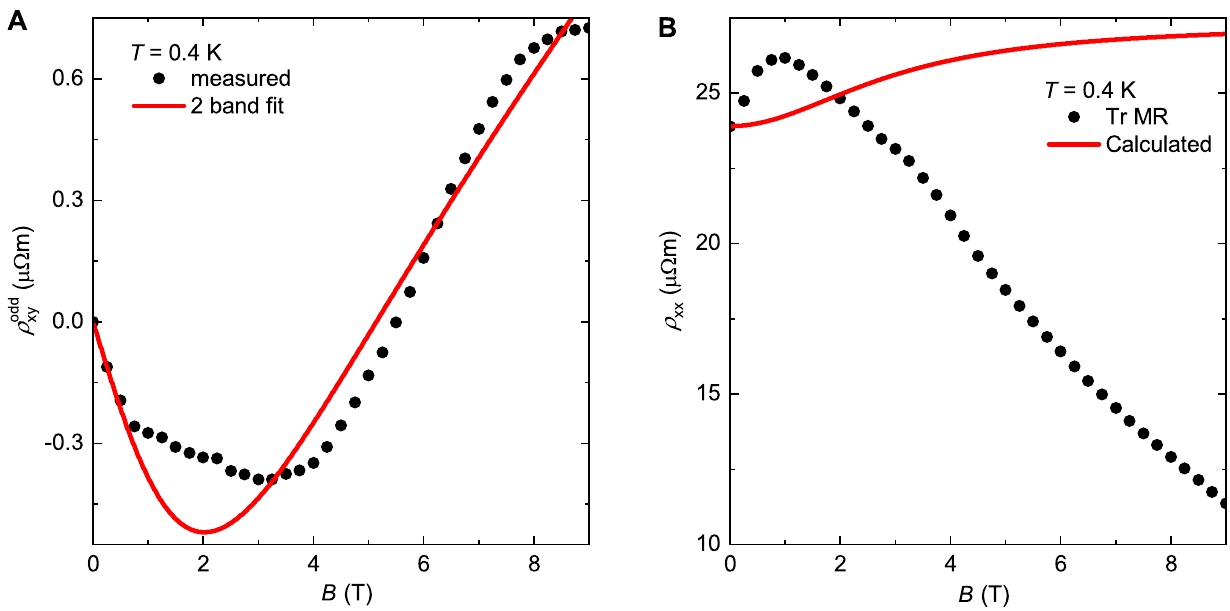}

\caption{\label{SMfig3} {\bf Two-band analysis of magnetotransport data of Ce$_3$Bi$_4$Pd$_3$.}
({\bf A}) Odd-in-field Hall resistivity at 0.4\,K (from Fig.\,\ref{Fig3}E of the
main part) with best fit of a two-band Drude model (red line, see text). ({\bf
B}) Transverse magnetoresistivity at 0.4\,K with corresponding curve calculated
from the parameters obtained in ({\bf A}), showing that a two-band model cannot
account for the experimental curves.}
\end{center}
\end{figure}

\noindent A nonlinear (transverse) magnetoresistance and (odd-in-field) Hall
resistivity, as observed for Ce$_3$Bi$_4$Pd$_3$ at low temperatures (see
Figs.\,\ref{Fig1}E and \ref{Fig3}E of the main part, respectively) might, {\em a
priori}, also result from multiple electronic bands contributing to the Fermi
surface. In the typically considered two-band case, a key requirement for strong
magnetotransport nonlinearities is a large difference in the carrier
concentration and mobility of the two bands (e.g., low-mobility majority
carriers and high-mobility minority carriers). To examine whether the
experimentally observed dependences, attributed to the odd-in-field anomalous
Hall effect (Fig.\,\ref{Fig3}E of the main part), might alternatively be
produced by two-band effects, we performed a two-band analysis using a recently
established robust analysis scheme \cite{Egu19.1}.

We use the $\rho_{xy}^{\rm{odd}}(B)$ data at the lowest temperature of 0.4\,K
(Fig.\,\ref{Fig3}E of the main part), where the nonlinearity is largest. Though
the obtained best fit cannot fully account for the field dependence, it
reproduces the overall shape of $\rho_{xy}^{\rm{odd}}(B)$ (Fig.\,\ref{SMfig3}A).
The obtained charge carrier concentrations and mobilities are $n_1 = 3.39 \times
10^{19}$\,cm$^{-3}$ and $\mu_1 = 105$\,cm$^2$/Vs for the (hole) majority
carriers and $n_2 = 1.24 \times 10^{17}$\,cm$^{-3}$ and $\mu_2 =
-3820$\,cm$^2$/Vs for the (electron) minority carriers. In a second step, we
calculate the transverse magnetoresistance for these parameters (red curve in 
Fig.\,\ref{SMfig3}B). It completely fails to describe the corresponding data. In
fact, no negative magnetoresistance can arise in this setting. Thus, we conclude
that two-band effects are not the cause of the observed nonlinearities in the
low-temperature magnetotransport data.

\subsection{Hall effect from odd-in-momentum Berry curvature: \SBP{Conceptual basis and data}}\label{nonlinearRH}

\noindent Here we provide supplementary information on the Hall effect
contribution in Ce$_3$Bi$_4$Pd$_3$ that is due to this (noncentrosymmetric)
material's odd-in-$k$ Berry curvature $\Omega_z^{\rm{odd}}(k)$ and leads to the
observed spontaneous Hall effect as well as the even-in-magnetic field Hall
component. The corresponding Hall conductivity $\sigma_{xy}$ (see Eqn.\,\ref{Eq1} of
the main part) depends on the out-of-equilibrium distribution function $f(k)$
under a driving electric field $\mathcal{E}_x$ and will thus be field dependent,
i.e., $\sigma_{xy}=\sigma_{xy}(\mathcal{E}_x)$. Thus, nonlinearities in terms of
$\mathcal{E}_x$ should appear.

In the perturbative treatment of Sodemann and Fu \cite{Sod15.1}, $\sigma_{xy}$
is linear in field, \vspace{-0.4cm}

\begin{equation}
\sigma_{xy}(\mathcal{E}_x) = \sigma_0\cdot\mathcal{E}_x \;\; ,
\end{equation}
\noindent where $\sigma_0$ is a constant, leading to a Hall current
density\vspace{-0.4cm}

\begin{equation}
j_{y} = \sigma_{xy}(\mathcal{E}_x)\cdot \mathcal{E}_x = \sigma_0\cdot\mathcal{E}_x^2 \label{Hallcurrent}
\end{equation}
\noindent that is quadratic in field. Under AC ($1\omega$) excitation, this Hall
current density appears as a $0\omega$ and $2\omega$ signal \vspace{-0.4cm}

\begin{equation}
j_{y}^{0\omega} = \sigma_0\cdot\mathcal{E}_x^{1\omega}(\mathcal{E}_x^{1\omega})^{\ast} \;\; \mbox{and} \;\; j_{y}^{2\omega} = \sigma_0\cdot(\mathcal{E}_x^{1\omega})^2 \;\; . \label{Eqn_jHall}
\end{equation}
\noindent In our experiments, we drive an electrical current $I_x$ through the
sample and measure a transverse voltage $V_{xy}$, and thus rewrite
Eqn.\,\ref{Hallcurrent} as \vspace{-0.4cm}

\begin{equation}
V_{xy} = \alpha I_x^2 \;\; .
\end{equation}
\noindent This square-in-current Hall voltage should appear in both DC and AC
experiments and we introduce the following nomenclature to distinguish the
different effects: \vspace{-0.4cm}

\begin{equation}
V_{xy}^{\rm{DC}} = \alpha^{\rm{DC}}\cdot(I_x^{\rm{DC}})^2 \;\; , \;\; V_{xy}^{0\omega} = \alpha^{0\omega}\cdot(I_x^{1\omega})^2 \;\; , \;\; V_{xy}^{2\omega} = \alpha^{2\omega}\cdot(I_x^{1\omega})^2 \;\; .
\label{eq_defs}
\end{equation}
\noindent These expressions are related to the corresponding Hall current
densities by $\mathcal{E}_x^{\rm{DC},1\omega} = \rho_{xx} j_x^{\rm{DC},1\omega}
= \rho_{xx} I_x^{\rm{DC},1\omega}/(w t)$ and $j_{y}^{\rm{DC},0\omega,2\omega} = 
\sigma_{yy}\mathcal{E}_{y}^{\rm{DC},0\omega,2\omega}= 
\sigma_{yy}V_{y}^{\rm{DC},0\omega,2\omega}/l_y$. The latter equation represents a current density that is opposite in sign to the Hall current density of Eqn.\,\ref{Hallcurrent}, thus fulfilling the open-circuit condition $j_{y} = \sigma_{xy}\cdot \mathcal{E}_x + \sigma_{yy}\cdot \mathcal{E}_y = 0$, i.e., no net current flows in the $y$ direction. Thus \vspace{-0.4cm}

\begin{equation}
\sigma_0 = \alpha \frac{1}{\rho_{xx}^3}\frac{w^2t^2}{l_y} \;\; ,
\end{equation}
\noindent where we have used $\rho_{xx} = 1/\sigma_{xx} = 1/\sigma_{yy}$, which
is exact for cubic symmetry, and where $w$, $t$, and $l_y$ are the sample width,
thickness, and transverse Hall contact distance, respectively. We will see that
while we indeed observe these three contributions, other (linear-in-current, but
still spontaneous or even-in-magnetic field) Hall voltage terms appear in
addition. As discussed in Sect.\,\ref{SINLH}, we attribute them to the tilted
Weyl-Kondo semimetal nature of Ce$_3$Bi$_4$Pd$_3$, which places the material in
the fully nonequilibrium electric field regime already at the smallest fields we
have applied.

First, we discuss the DC experiments. The longitudinal voltage $V_{xx}$ is
linear in $I$ (defined as applied along $x$). \SBP{This simple Ohmic behaviour demonstrates that the applied electric field has no appreciable effect on the material's bandstructure. By contrast,} the spontaneous Hall voltage
$V_{xy}$ is nonlinear in $I$, and well described as the sum of a linear
($R_{xy}I$) and quadratic ($\alpha^{\rm{DC}} I^2$) term (Fig.\,\ref{dc}A).
\begin{figure}[t!]
\centering
\includegraphics*[width=0.8\textwidth]{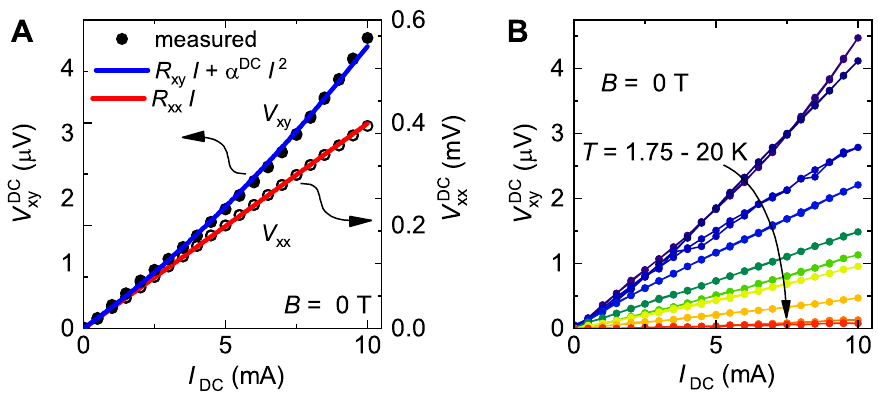}

\caption{\label{dc} {\bf Nonlinearity of the spontaneous Hall effect in DC
transport.} ({\bf A}) Current-voltage characteristics in zero magnetic field for
the transverse (full symbols, left axis) and  longitudinal (open symbols, right axis) voltage of
Ce$_3$Bi$_4$Pd$_3$ at 1.75\,K. The nonlinearity in the Hall response is
quantified by $\alpha^{\rm DC}$. ({\bf B}) Temperature evolution of the
current-voltage characteristic of the Hall component in zero magnetic field.}
\end{figure}
Both emerge only as Kondo coherence sets in at low temperatures (Fig.\,\ref{dc}B
and Fig.\,\ref{Fig2}F of the main part). Only the quadratic-in-field (or
current) response was predicted in ref\,\cite{Sod15.1}. The joint appearance
of the linear-in-field term strongly suggests that both phenomena have a common
origin (see Sect.\,\ref{SINLH} for its discussion). That the emergence of a spontaneous Hall voltage is due to a thermodynamically as opposed to microscopically broken TRS (as in magnetic systems) is further discussed in Sect.\,\ref{TRS_breaking}.

Next, we address nonlinearities in our AC experiments. We observe both a
$2\omega$ and a $0\omega$ spontaneous Hall voltage in response to a $1\omega$
current excitation (Fig.\,\ref{Fig2}D,\,E of the main part and Fig.\,\ref{ac1omega}A).
\begin{figure}[h!]
\centering
\includegraphics*[width=0.8\textwidth]{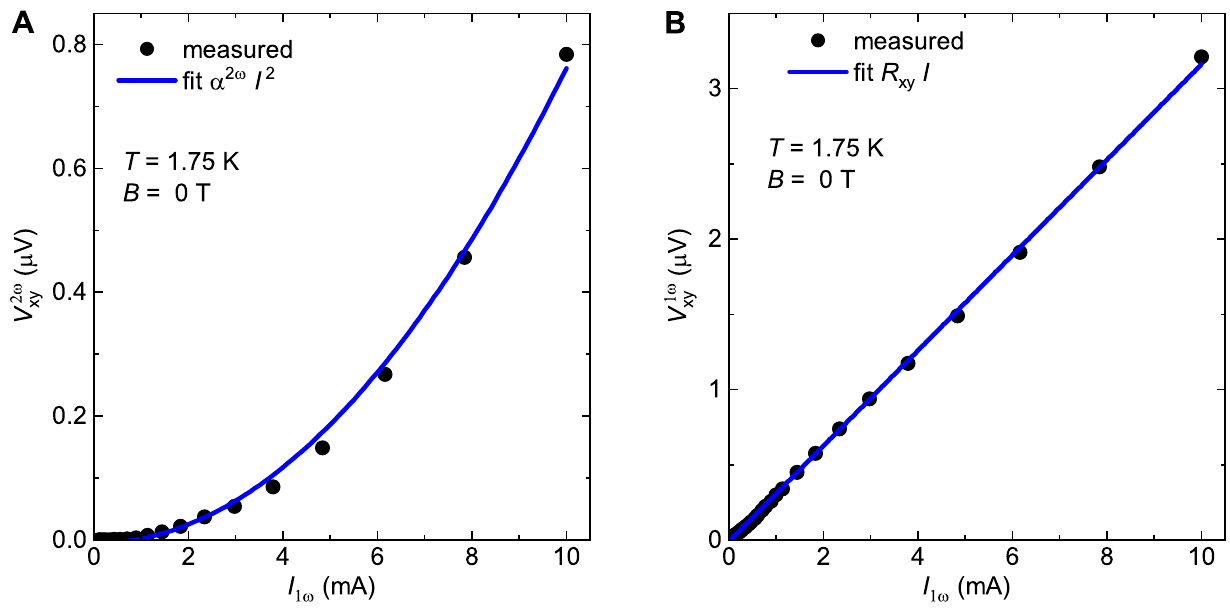}

\caption{\label{ac1omega} {\bf Nonlinearity of the spontaneous Hall effect in AC transport.} ({\bf A}) Second harmonic Hall voltage of Ce$_3$Bi$_4$Pd$_3$ in response to an AC excitation in zero magnetic field and at 1.75\,K. The blue curve is a quadratic fit to the data, with the fitting parameter $\alpha^{2\omega}$. ({\bf B}) First harmonic, linear-in-current Hall voltage measured under the same condition as in panel ({\bf A}). The blue line is a linear fit to the data, with the slope $R_{\text{xy}}$. The measurement was done on the same sample as in Fig.\,\ref{dc}.}
\end{figure}
Both emerge only upon entering the Kondo coherent regime (Fig.\,\ref{Fig2}F of
the main part). In addition to these two contributions, which were predicted in
ref\,\cite{Sod15.1}, we also find a robust $1\omega$ spontaneous Hall signal,
that closely follows the linear-in-current signal of our DC experiment
(Fig.\,\ref{Fig2}B of the main part and Fig.\,\ref{ac1omega}B). Both display a
\SBP{pronounced} linear $\sigma_{\rm{xy}}$ vs $\sigma_{\rm{xx}}$ relationship in the
Kondo coherent regime (grey shaded area in Fig.\,\ref{Fig2}B of the main part,
temperature is an implicit parameter). In terms of absolute values, this is the
dominant signal; it is of the order of the 3D conductivity quantum $e^2/(h a)$,
where $a$ is the lattice parameter. The absolute values of the different
contributions are given in Table~\ref{SMtable}.
\begin{table}[h!]
\caption{\label{SMtable} Absolute values of the spontaneous Hall response of single crystal S2, with current approximately along $[111]$, as defined in Fig.\,\ref{Fig2} of the main part.}
\begin{tabular}{|c|ccccc|}
 \hline
& $2\omega$ & $0\omega$ & DC-$I^2$ & $1\omega$ & DC-$I$\\
\hline
$T_{\rm H}$ (K) & 3.8 & 4.6 & 4.3 & 4.8 & 4.9 \\
$\alpha^{\rm max}$ (10$^{-3}$\,$\mu$V/mA$^2$) & 7.67      &  10.8     & 14.3 & & \\
$\rho_{xy}^{\rm max}$ ($\mu\Omega$m) & & & & 0.21 & 0.18\\
 \hline
\end{tabular}
\end{table}

We stress that the spontaneous Hall response is robust in all channels
(linear-in-$I$ and quadratic-in-$I$ Hall voltage in DC experiments, and
$0\omega$, $1\omega$, and $2\omega$ Hall voltage in AC experiments). Among the
many samples we have measured, not a single one did not show a spontaneous Hall
effect. For the three samples we have studied in depth (three single crystals
from different growth batches, with current excitation approximately along
$[103]$, $[100]$, and $[111]$, respectively), all features are very similar
(Fig.\,\ref{reproduce}). In terms of absolute values, $\tan\Theta_{\rm H}$
(in the linear-in-$I$ response for crystal S1 and in
the $1\omega$ response for crystals S2 and S3) is 0.19, 0.52, and 0.17,
respectively.

\begin{figure}[t!]
\centering
\includegraphics*[width=0.95\textwidth]{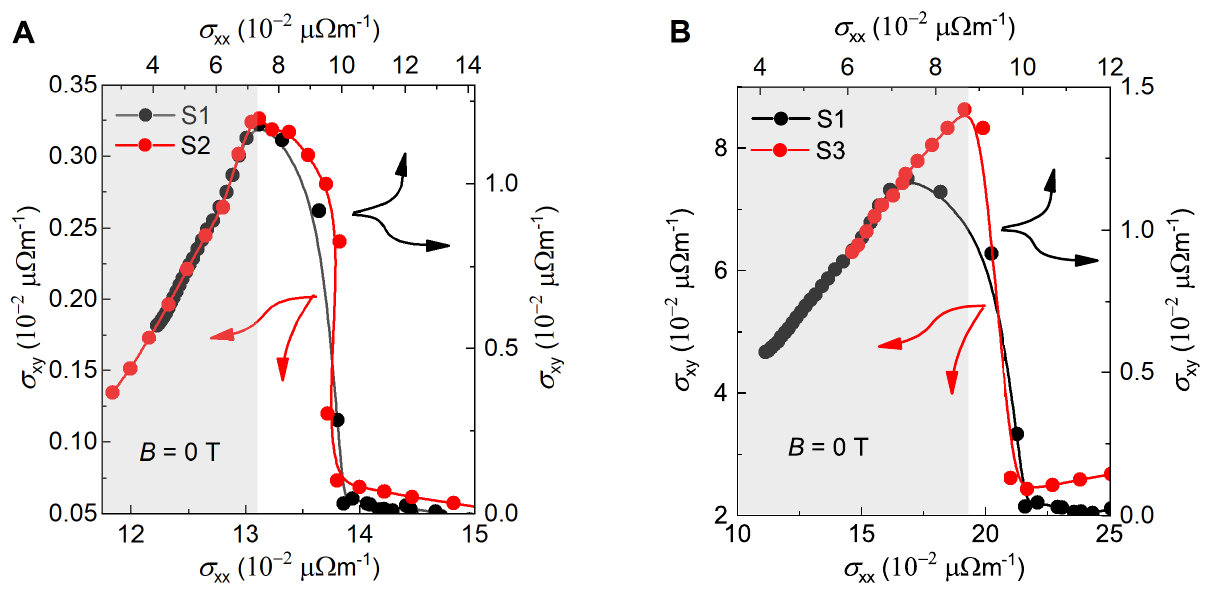}

\caption{\label{reproduce} {\bf Reproducibility of the spontaneous Hall
effect.} Linear-in-current spontaneous Hall signal, shown as $\sigma_{\rm{xy}}$
vs $\sigma_{\rm{xx}}$, where temperature is an implicit parameter, for samples
from three different growth batches and orientations (current approximately
along $[103]$, $[111]$, and $[100]$ for crystals S1, S2, and S3, respectively).
S1 was measured with DC current in quasi-AC mode, S2 and S3 are $1\omega$
signals from AC experiments. ({\bf A}) Comparison of crystals S1 and S2. ({\bf B})
Comparison of crystals S1 and S3. The grey shading indicates the Kondo coherent
regime, where $\sigma_{\rm{xy}}$ \SBP{is linear}
in $\sigma_{\rm{xx}}$.}
\end{figure}

Finally, we comment on effects in a finite applied magnetic field $B$. Here,
Hall contributions associated with the odd-in-$k$ Berry curvature are detected
as even-in-$B$ contributions, which is how we separate them from other (normal,
anomalous, two-band ...) contributions, which are all odd in $B$. Again, such
even-in-$B$ contributions appear in all channels (see Fig.\,\ref{Fig2}D,\,E and
Fig.\,\ref{Fig3}C,\,D) and represent the same physics discussed in the main
part. The only role the magnetic field plays here is to successively suppress
the odd-in-$B$ Hall contribution, similar to what happens when enhancing the
temperature above the Kondo coherence temperature.

\subsection{Berry curvature-driven Hall effect and macroscopic TRS breaking}\label{TRS_breaking}

\noindent The Berry curvature-driven Hall effect  of a noncentrosymmetric
material with (microscopically) preserved TRS (Eqn.\,\ref{Eq1}), though measured
under zero magnetic field, is a nonlinear response effect to an applied electric
field $\mathcal{E}_x$. As such it is inherently dissipative, causing TRS
breaking on the thermodynamic level. This is in contrast to the linear response
regime in which the normal (Lorentz force-driven) Hall effect can be considered.

The nonlinear response is determined by the distribution function in the
presence of the external drive, which breaks TRS even if the system in
equilibrium obeys TRS. In fact, the corresponding Hall conductivity
$\sigma_{xy}$ is defined only in the presence of an electric field, and hence in
a state that produces entropy (at the rate $\dot{S} = j_x \mathcal{E}_x$) and
breaks TRS thermodynamically. In other words, in the limit of zero entropy
production, that is for $\mathcal{E}_x \rightarrow 0$, also $\sigma_{xy}
\rightarrow 0$.

This is in contrast to electrical transport in the framework of linear response theory, where the transport coefficients are expressed in terms of correlation functions in the absence of a driving field. As a consequence these correlation functions reflect the TRS of the system in equilibrium. Thus, even though an applied electric field by itself breaks TRS on the thermodynamic level, the linear response transport coefficients are still well defined when the entropy production is vanishing, i.e., for $\mathcal{E}_x \rightarrow 0$ the Hall conductivity remains finite.\vspace{0.5cm}

\subsection{Odd-in-current contribution to Berry-curvature driven Hall voltage}\label{OddInCurrent}

\noindent The Berry curvature-driven Hall current density in an IS breaking but TRS preserving material is given by\vspace{-0.4cm}

\begin{equation}
j_{y} = \frac{e^2}{\hbar} \int \frac{d^3k}{(2 \pi)^3}  f(\bm{k}) \underbrace{\Omega^{\rm{odd}}_z(\bm{k})\mathcal{E}_x}_{v_y}
\label{Eqn_SHE_sigma}
\end{equation}
\noindent where $f(\bm{k})$ is the out-of-equilibrium distribution function associated with an applied electric field $\mathcal{E}_x$ and $v_y$ is the anomalous velocity. The specific dependence of the Hall voltage on the current, $V_{xy}(I_x)$, associated with Eqn.\,\ref{Eqn_SHE_sigma} depends on the form of $f(\bm{k})$. In a linearized Boltzmann approximation (as used in refs\,\cite{Sod15.1,Zha18.1}), $f(\bm{k})$ corresponds to the equilibrium (Fermi-Dirac) distribution function $f_0(\bm{k})$ rigidly shifted in the direction of $\mathcal{E}_x$, that is $f(\bm{k}) = f_0(\bm{k}) + \partial f_0/\partial k_x \cdot \delta k_x$, with $\delta k_x = (e\tau/\hbar)\cdot\mathcal{E}_x$. In this case, Eqn.\,\ref{Eqn_SHE_sigma} simplifies to \cite{Sod15.1} \vspace{-0.4cm}

\begin{equation}
j^{(2)}_y=\frac{e^3\tau \mathcal{E}_x^2}{\hbar^2} \underbrace{\int \frac{d^3k}{(2 \pi)^3} \frac{\partial f_0(\bm{k})}{\partial k_x}\Omega^{\rm{odd}}_z(\bm{k})}_{D_{zx}} \; \; ,
\label{Eqn_SHE_Sod}
\end{equation}
\noindent where $D_{zx}$ is the Berry curvature dipole and the superscript (2) makes it explicit that this is a 2nd order effect in $\mathcal{E}_x$, i.e., $j_{y}$ is quadratic and thus even in $\mathcal{E}_x$. However, this picture fails to consider the boundary condition of a Hall effect experiment, namely that no net current can flow in the $y$-direction (see Sect.\,\ref{nonlinearRH}, text below Eqn.\,\ref{eq_defs}). To fulfill this condition, the above current $j_{y}^{(2)}$ must be compensated by an equal current of opposite sign, that can only be generated by a shift of the Fermi surface in the $k_y$ direction by $\delta k_y =  (e\tau/\hbar) \cdot \mathcal{E}^{(2)}_y$, with $\mathcal{E}^{(2)}_y = -\rho_{xx} j_{y}^{(2)}$. In a second step, this shift then leads to an additional Berry curvature-driven Hall current density $j^{(3)}_y$, by populating states with finite anomalous velocity $v_y(\bm{k})$.
 
To be explicit, the (extra) shift of $f(\bm{k})$ by $\delta k_y$,\vspace{-0.4cm}

\begin{equation}
f(\bm{k} + \delta \bm{k}_y) = f^0(\bm{k}) + \delta k_x \frac{\partial f^0(\bm{k})}{\partial k_x} + \delta k_y \frac{\partial f^0(\bm{k})}{\partial k_y} \; \; ,
\label{Eqn_FermiShift1}
\end{equation}
\noindent results in an extra contribution (3rd term) which, through
Eqn.\,\ref{Eqn_SHE_sigma} and with $\delta k_y = - (e^4\tau^2 \rho_{xx}/\hbar^3) \cdot  D_{zx}\mathcal{E}_x^2$, drives the (3rd order) Hall current density \vspace{-0.4cm}

\begin{equation}
j^{(3)}_y = - (\frac{e^3 \tau}{\hbar^2})^2 \rho_{xx} D_{zx}D_{zy} \mathcal{E}_x^3 \; \; ,
\label{Eqn_SHE_Sod2}
\end{equation}
\noindent where $D_{zy} =  \int d^3k/(2\pi)^3 (\partial f_0(\bm{k})/\partial k_y) \Omega^{\rm{odd}}_z(\bm{k})$ is the corresponding Berry curvature dipole.
Combining Eqns.\,\ref{Eqn_SHE_Sod} and \ref{Eqn_SHE_Sod2}, the Hall voltage $V_{xy}$ will be the sum of both even-in-$I_x$ and odd-in-$I_x$ contributions.\vspace{-0.3cm}

\subsection{Trivial Hall contribution from crystal symmetry}\label{trivialRH}

\noindent Finally, we note that in systems with low enough crystalline
symmetry (lower than cubic), the resistivity tensor cannot be represented by a
simple scalar. As a consequence, a Hall-like linear-in-$\mathcal{E}$ ($1\omega$)
signal might appear if the electric field $\mathcal{E}$ is applied off-axis with
respect to the main crystallographic directions.

However, Ce$_3$Bi$_4$Pd$_3$ is cubic and thus no such (trivial) perpendicular
component can arise. This can be readily understood: For a cubic system, the
three eigenvalues of  the conductivity matrix are equal to each other; in other
words, the conductivity matrix is a constant multiplied by a unit matrix. In
this case, regardless of the direction along which the drive electric field is
applied, the induced current from this mechanism will always be along the
direction of the applied field, and there is no Hall response. Note that the
Weyl-Kondo phase in equilibrium preserves the crystalline symmetry.

One can make this rather transparent point more formally by perturbatively
solving the Boltzmann equation for the non-equilibrium distribution function $f$
and see that this ``trivial" mechanism would in general yield $1\omega$,
$3\omega$ {\it etc.} contributions to the Hall response.  However, for the cubic
system, the contributions vanish. We can illustrate the point by considering
the  leading order (responsible for any $1\omega$ response)\vspace{-0.4cm}

\begin{eqnarray}
&& f_1^\omega = \frac{e\tau \mathcal{E}_a \partial_a f_0}{1+ i \omega \tau} \;\; ,
\nonumber
\end{eqnarray}
where  $f_0$ is the equilibrium distribution, and the repeated indices mean
summation over. In turn, the $1\omega$ anisotropic electric currents induced by
the ``normal" velocity term is \vspace{-0.4cm}

\begin{eqnarray}
j_a^\omega &= & -e\int_k \partial_a \epsilon_k f_1^\omega \;\; .
\end{eqnarray}
Combining these equations, we find the normal-velocity induced $1\omega$ current
to be \vspace{-0.4cm}

\begin{eqnarray}
j_{a}^\omega =  -\frac{e^2 \tau}{1+i\omega \tau} \int_k \left( \partial_a \epsilon_k\right) \left( \partial_\mu f_0 \right) \mathcal{E}_\mu \;\; .
\label{Eq:ja1} 
\end{eqnarray}
An electric field $\mathbf{\mathcal{E}}$ applied in an arbitrary direction can
be expressed as the combination of its components along the three principle
axes, \textit{i.e.}, $\mathbf{\mathcal{E}}= \mathcal{E}_x \hat{x} + 
\mathcal{E}_y \hat{y} + \mathcal{E}_z \hat{z}$. According to
Eqn.~\eqref{Eq:ja1},  $\mathcal{E}_x$ only induces a current along $\hat{x}$
(otherwise, the integral is over an odd function and vanishes): $j_{x}^\omega =
\sigma_{xx} \mathcal{E}_x$. Likewise, $\mathcal{E}_y$ induces a current along
$\hat{y}$; and $\mathcal{E}_z$ induces a current along $\hat{z}$. Moreover,
$\sigma_{xx}=\sigma_{yy}=\sigma_{zz}$. Thus, the induced current must be along
the  electric field, regardless of the direction the electric field is applied,
and there is no Hall response.\vspace{-0.3cm}

\subsection{Imprint of Berry curvature-driven Hall effect on electrical resistivity}\label{SI_RxxJump}

\noindent We observe that the onset of spontaneous Hall effect with decreasing temperature leaves an imprint also on the (longitudinal) electrical resistivity (Fig.\,\ref{SMfig2}). This effect is understood as follows: Firstly, there is a trivial source, a transverse direction misalignment of the electrical resistance contacts (Fig.\,\ref{SMfig_ContactSketch}), that is explained in detail in Sect.\,\ref{misaligne}.
Secondly, there is an intrinsic source, which is discussed in what follows.

The appearance of a Hall voltage is associated with a reconstruction of the current path in the sample, charge carriers being accumulated on one side of the sample (hence the Hall voltage). This will not affect the longitudinal resistivity appreciably if the tangent of the Hall angle, $\tan{\Theta_{\rm H}} = \mathcal{E}_y/\mathcal{E}_x = \sigma_{xy}/\sigma_{xx}$, is small. However, with the giant values we observe ($\tan{\Theta_{\rm H}}$ up to 0.5, see main text), we are not in this limit and hence do observe a nonnegligible effect. To be explicit, the resistivity $\rho_{xx}$ depends on $\tan\Theta_{\rm H}$ as \vspace{-0.4cm}

\begin{equation}
\rho_{xx} = \frac{\sigma_{xx}}{\sigma_{xx}^2 + \sigma_{xy}^2} = \frac{1}{\sigma_{xx}}\frac{1}{1+\tan^2\Theta_{\rm H}}
\label{Eqn_Rxx_drop}
\end{equation}

Thus, as $\tan\Theta_{\rm H}$ increases upon the onset of the spontaneous Hall
effect, $\rho_{xx}$ should drop. To test this scenario, we plot the relative
change of resistivity across the onset of the spontaneous Hall effect as a
function of temperature (Fig.\,\ref{Rxx_jump}A-C). Indeed, we observe the
expected drop for all three samples studied. Moreover, we observe that the relative drop size increases with $\tan\Theta_{\rm H}$, and that the
magnitude and functional form of the dependence (Eqn.\,\ref{Eqn_Rxx_drop}) is in
overall agreement with the data. This provides strong evidence that the drop in
$R_{xx}$ is indeed due to this intrinsic effect.
\begin{figure}[t!]
\centering
\includegraphics*[width=0.75\textwidth]{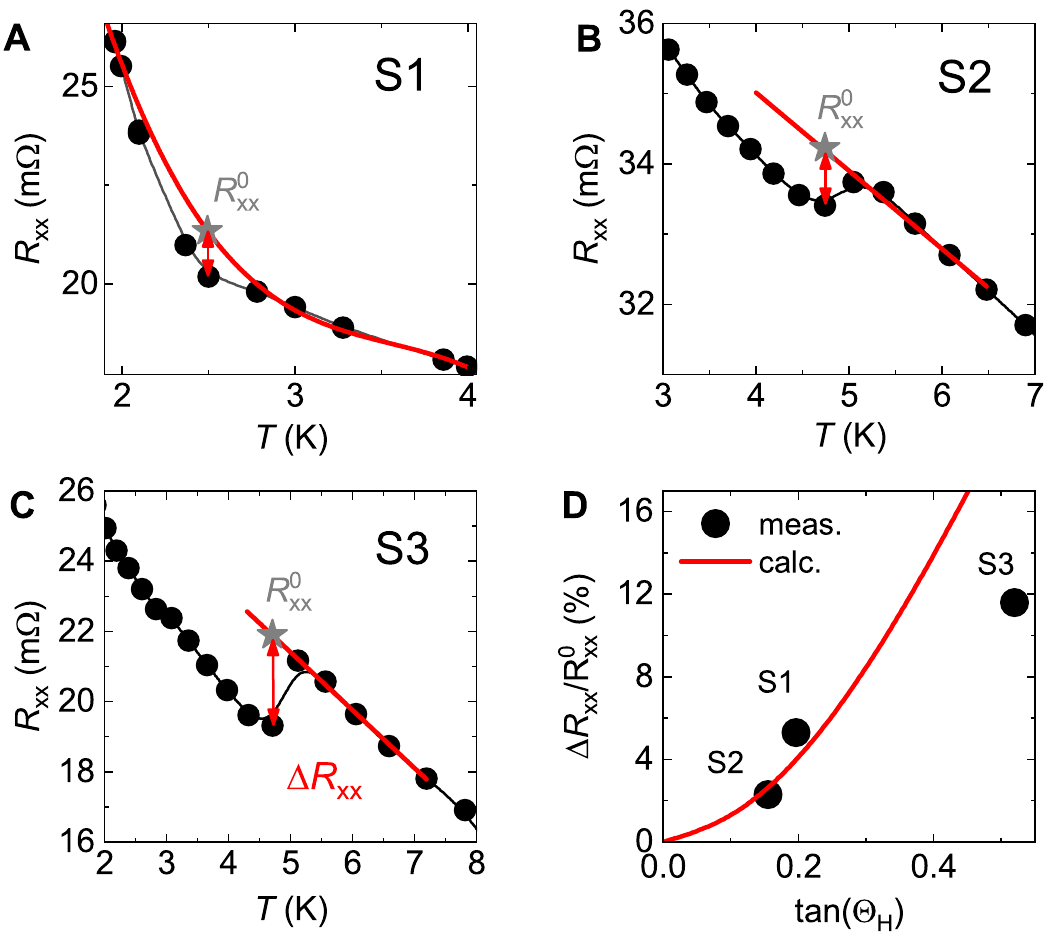}

\caption{\label{Rxx_jump} {\bf Resistance change due to large Hall angle.} ({\bf
A}-{\bf C}) Longitudinal electrical resistance $R_{xx}$, corrected for contact
misalignment (Fig.\,\ref{SMfig_ContactSketch}), for samples S1, S2, and S3. With decreasing
temperature, a drop of $R_{xx}$ is observed at the onset of the spontaneous Hall
effect. For a rough estimate of its magnitude, $R_{xx}(T)$ above the onset
temperature is extrapolated to lower $T$ and the difference $\Delta R_{xx}$ is
read off the data as indicated (red double arrow). ({\bf D}) Relative resistance
change $\Delta R_{xx}/R_{xx}^0$ of the three sample (symbols) vs the tangent of
the Hall angle (determined at low temperatures where the spontaneous Hall effect is fully established), defined as $\tan{\Theta_{\rm H}} =
\partial\sigma_{xy}/\partial\sigma_{xx}$ (see Fig.\,\ref{reproduce}). Clearly, $\Delta
R_{xx}/R_{xx}^0$ increases with $\tan{\Theta_{\rm H}}$. The red line shows the
expectation from Eqn.\,\ref{Eqn_Rxx_drop}, $\Delta\rho_{xx}/{\rho_{xx}^0} =
(\rho_{xx}^0-\rho_{xx})/\rho_{xx}^0 = 1- 1/(1+\tan^2\Theta_{\rm H})$, with
no adjustable parameter. The overall agreement, both in terms of magnitude and
shape, further underpins that the observed signatures in $R_{xx}$ are intrinsic
and due to the giant magnitude of the Hall angle.}
\end{figure}

That the Hall angle is different for the different samples is, most likely, due to the different current directions with respect to the crystal axes. Preliminary results on a single crystal contacted several times to achieve different current directions reveal a sinusoidal variation of the magnitude of the spontaneous Hall resistivity with angle. Such a dependence is expected for a Berry-curvature driven Hall effect.

\section{$\mu$SR, magnetization, and specific heat}\label{SImuSRCp}

\subsection{Analysis of $\mu$SR data}\label{SImuSR}

\noindent Due to the large muon gyromagnetic ratio and the availability of 100\%
spinpolarized muon beams, ZF-$\mu$SR is one of the most sensitive probes for
detecting small spontaneous magnetic fields. Consequently, the technique has
been successfully used to determine the occurrence (or the absence) of TRS
breaking in many different materials \cite{Luk98.1,Hil09.1,Sha18.1}. If TRS is
broken, the onset of tiny spontaneous currents gives rise to associated (weak)
magnetic fields, detected by ZF-$\mu$SR as an increase in the muon spin
relaxation rate.

In nonmagnetic materials in zero field, the relaxation is typically dominated 
by the randomly oriented nuclear moments (in our case by the $4.11\mu_{\rm{N}}$
nuclear moment of $^{209}$Bi), which can be described by the Gaussian
Kubo-Toyabe relaxation function \cite{Yao11.1} \vspace{-0.4cm}

\begin{equation}
G_\mathrm{KT} = \frac{1}{3} + \frac{2}{3}(1
-\sigma^{2}t^{2})\,\mathrm{e}^{-\frac{\sigma^{2}t^{2}}{2}} \;\; .
\end{equation}
\noindent A possible electronic contribution to the ZF-$\mu$SR spectra is
modeled by an additional Lorentzian relaxation, with the electronic relaxation
rate denoted by $\Lambda$. The total relaxation is then described by
\vspace{-0.4cm}

\begin{equation}
\label{eq:KT_and_electr}
A_\mathrm{ZF} = A_\mathrm{s} G_\mathrm{KT} \mathrm{e}^{-\Lambda t} + A_\mathrm{bg} \;\; ,
\end{equation}
\noindent where $A_\mathrm{s}$ and $A_\mathrm{bg}$ are the sample- and
background-related asymmetries, respectively, the latter being about 15\% of the
former in our case.

\begin{figure}[!t]
\centering
\includegraphics[width=0.5\columnwidth]{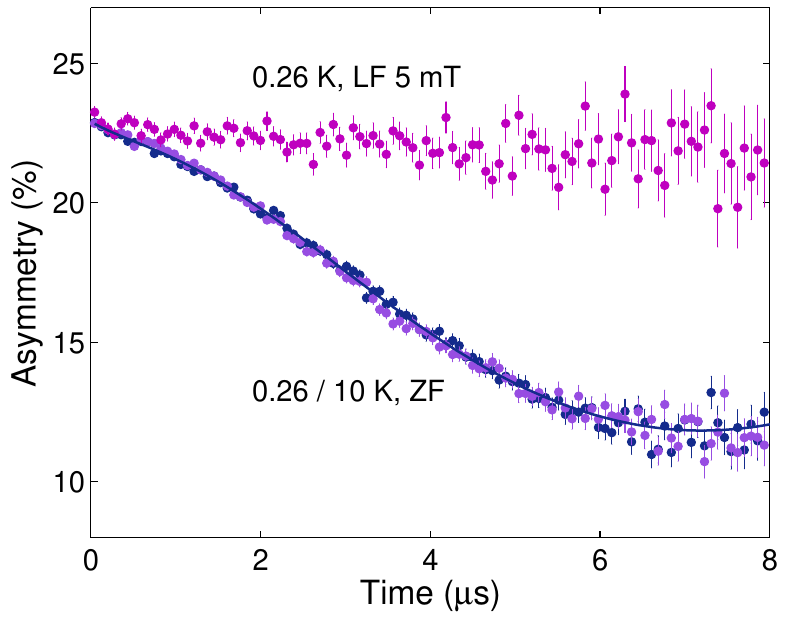}

\caption{\label{fig:ZF_LF_muSR} \textbf{ZF-$\mu$SR detects no TRS breaking in
Ce$_3$Bi$_4$Pd$_3$.} ZF-$\mu$SR spectra at 0.26 and 10\,K, and 5\,mT
longitudinal-field (LF) spectrum at 0.26\,K. The collapse of the two ZF spectra
is direct evidence that no magnetization develops in this temperature range. The
weak muon spin depolarization is mostly of nuclear origin, as confirmed by its
prompt recovery in a very small field. The solid line is a fit to the spectra by
means of Eqn.\,\ref{eq:KT_and_electr}.}
\end{figure}

The ZF-$\mu$SR data of Ce$_3$Bi$_4$Pd$_3$ taken at 10 and 0.26\,K essentially
collapse on top of each other (Fig.\,\ref{fig:ZF_LF_muSR}), providing direct
evidence that no magnetization develops in this temperature range. Fits to all
data sets were performed using the \texttt{musrfit} suite \cite{Sut12.1}, with
the raw error bars reflecting the data counting statistics and the fit parameter
uncertainties resulting from standard error propagation methods. The resulting
fit parameters reveal that the depolarization is mostly of nuclear origin, which
is further confirmed by the prompt recovery of the signal in a very small
longitudinal magnetic field (Fig.\,\ref{fig:ZF_LF_muSR}). The electronic
relaxation rate is very small and temperature independent within the error bars
(see Fig.\,\ref{Fig2}C of the main part). This is unambiguous evidence that, in
the studied temperature range, TRS is preserved in Ce$_3$Bi$_4$Pd$_3$.\vspace{0.5cm}

\subsection{Magnetization}\label{SImagnetization}

\noindent As discussed above, ZF-$\mu$SR is the most sensitive probe to detect TRS breaking, which may for instance be caused by a magnetic phase transition (even if partial or due to a minority foreign phase), and has clearly ruled it out. Nevertheless, to corroborate this finding with a more common tool, we have performed field-sweep measurements of the magnetization at temperatures below the onset of the spontaneous Hall effect (Fig.\,\ref{Fig:SI_hysteresis}). Any even spurious ferromagnetic transition would lead to hysteretic behavior. Clearly, no hysteresis is observed, neither in sweeps up to high fields (up to 2.5\,T, panel A), nor in precise very low-field sweeps (up to 0.15\,T, panel B). In fact, the $M(H)$ curves in both field ranges show a smooth, linear behavior, as expected for a paramagnetic phase.

\begin{figure}[h!]
\centering
\includegraphics*[width=0.97\textwidth]{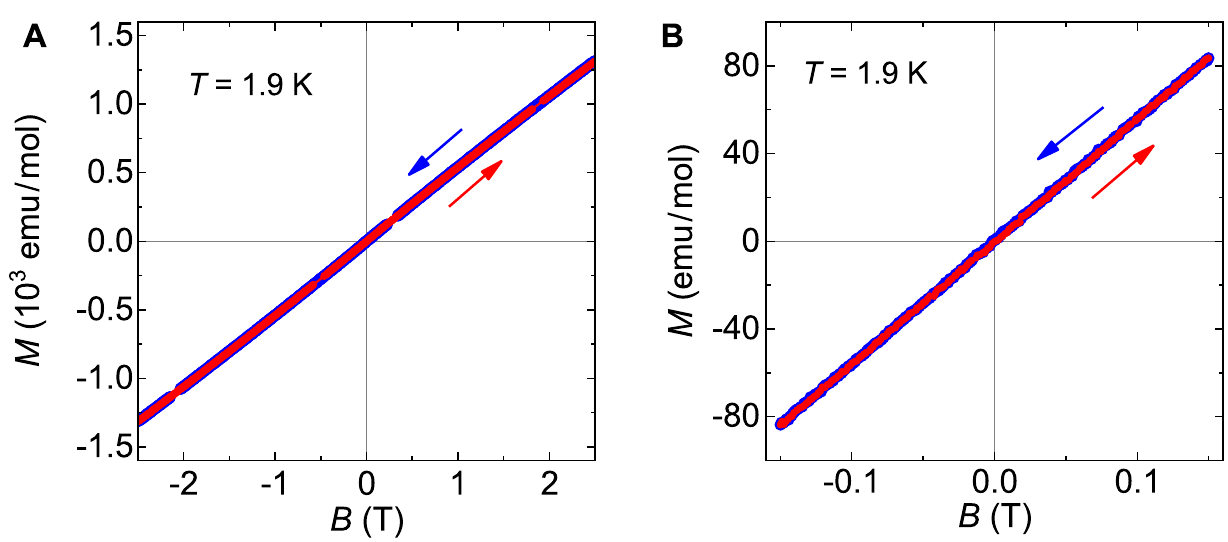}
\caption{\label{Fig:SI_hysteresis}{\bf Absence of hysteresis in the magnetization}. Magnetization loops of Ce$_3$Bi$_4$Pd$_3$ in fields up to 2.5\,T (left) and higher accuracy measurement in fields up to 150\,mT (right), both revealing the absence of any hysteretic behavior. The symbol sizes for up (red) and down (blue) sweeps are different for better visibility. The initial magnetization curve is contained in the up sweep.}
\end{figure}

\begin{figure}[h!]
\centering
\includegraphics[width=0.47\columnwidth]{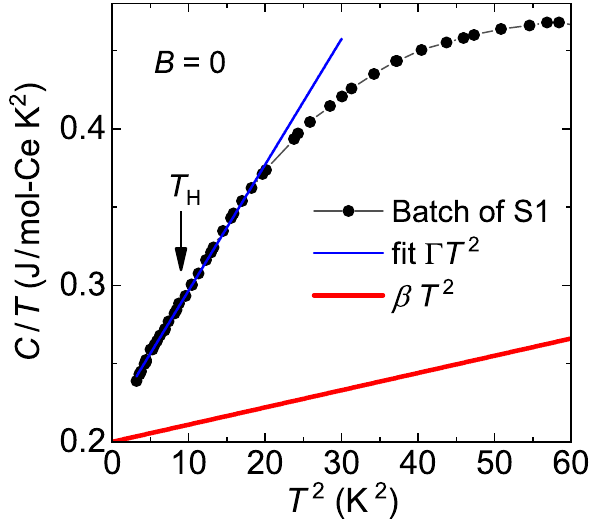}

\caption{\label{fig:Cp_S1} \textbf{Specific heat of Ce$_3$Bi$_4$Pd$_3$.} Specific heat in zero magnetic field, plotted as $C/T$ vs $T^2$ (black), for the batch of sample S1. The data follow a linear behavior (blue line) with a slope that overshoots the Debye phonon term (red line, offset for clarity, taken from ref\,\cite{Dzs17.1}), as expected for a Weyl-Kondo semimetal. No phase transition is discerned, neither at the onset of the spontaneous Hall effect ($T_{\rm H}$), nor elsewhere.}
\end{figure}

\subsection{Specific heat measurements}\label{SICp}

\noindent Specific heat measurements are the most direct tool to detect phase transitions, be it with or without TRS breaking. In Fig.\,\ref{fig:Cp_S1} we show the specific data for the batch of sample S1, taken at zero magnetic field and plotted as $C/T$ vs $T^2$. As seen previously \cite{Dzs17.1}, an enhanced $C/T = \Gamma T^2$ term [with $\Gamma = 0.008$\,J/(mol-Ce\,K$^4$), corresponding to a low quasiparticle velocity of about 1000\,m/s] is observed, evidencing very flat linearly-dispersing electronic bands, one of the hallmarks of a Weyl-Kondo semimetal \cite{Dzs17.1,Lai18.1}.

No phase transition anomaly is seen in the data. Most notably, at the onset temperature of the spontaneous Hall effect ($T_{\rm H}$), the data vary smoothly, following the $C/T$ vs $T^2$ law without any detectable deviation.

In summary, our ZF-$\mu$SR, magnetization, and specific heat data rule out that the observed spontaneous and even-in-magnetic field Hall effect seen in our Ce$_3$Bi$_4$Pd$_3$ samples is due to extrinsic effects.

\section{Density functional theory calculations}\label{SIDFT}

\noindent For the density functional theory (DFT) bandstructure calculations of
Ce$_3$Bi$_4$Pd$_4$ we used the all-electron augmented-plane-wave code WIEN2k
\cite{Bla18.1} with the Perdew, Burke, Ernzerhof (PBE) functional
\cite{Per96.1}. The experimentally determined structure \cite{Her08.2} was used
with space group $I\bar{4}3d$, $a = 10.052$\,\AA, and Ce, Pd, and Bi at the
$12a$, $12b$, and $16c$ (with $x=0.0839$) positions, respectively. As DFT cannot
treat many body effects such as the Kondo interaction of the Ce $4f$ electrons
with the conduction electrons derived from $s$, $p$, and $d$ orbitals, we use it
here to describe the conduction electron part only. For this purpose, we use the
open-core approximation, where one Ce $4f$ electron is included as
noninteracting core state, but the $4f$ basis functions are removed from the
bandstructure calculations and thus cannot hybridize with other orbitals. As
such, this approach is the {\em ab initio} counterpart of $H_c$ in the periodic
Anderson Hamiltonian of the Weyl-Kondo model (Eqn.\,\ref{PAM1}). 

We used atomic sphere radii of 3.0, 2.5, and 2.5 Bohr for Ce, Pd and Bi,
respectively, and a plane wave cutoff parameter of $R K_{\rm{max}}=8.5$, where
$R=2.5$ and $K_{\rm{max}}$ restricts the plane wave expansion. Inside the atomic
spheres an angular momentum expansion up to $l_{\rm{max}}=10$ is used, while for
the nonspherical matrix elements $l_{\rm{max}}$ is restricted to 6. The
$\bm{k}$-space integration was done with an $18 \times 18 \times 18$ $\bm{k}$
mesh. Spin-orbit coupling was included in a second variational procedure
\cite{Bla18.1}.

\begin{table}[!t]
\caption{\label{SMtable1} Coordinates of Weyl nodes in the $k_x$-$k_z$ plane of the Brillouin
zone of Ce$_3$Bi$_4$Pd$_3$, in units of $2\pi/a$, where $a=10.052$\,\AA, and
their energies with respect to the Fermi energy in eV. Due to the cubic symmetry
of the system, all coordinates occur in both signs and all $x$, $y$, $z$
permutations, leading to a multiplicity of 24.}
\begin{tabular}{| c | c c c | c |} 
 \hline
 Weyl node & $k_x$ & $k_y$ & $k_z$ & $E$\\ [0.5ex] 
 \hline\hline
 1 & 0.517 & 0 & 0.174 & -0.127\\ 
 \hline
 2 & 0.447 & 0 & 0.164 & -0.174\\
 \hline
 3 & 0.372 & 0 & 0.112 & -0.196\\
 \hline
 4 & 0.172 & 0 & 0.087 & -0.282\\
 \hline
\end{tabular}

\end{table}

To identify the Weyl nodes, we used a modified version of BerryPI \cite{Ahm13.1}
that calculates the Berry curvature of Wilson loops around candidate Weyl nodes
\cite{Wen15.1}. Within the $k_x$-$k_z$ plane of the Brillouin zone, we have
identified 4 Weyl nodes (see Fig.\,\ref{Fig4}A of the main part) with the
coordinates and energies given in Table~\ref{SMtable1}. All Weyl nodes are
situated at least 100\,meV away from the Fermi level. We expect that the Kondo
interaction will pin (part of) these Weyl nodes to the Fermi level, just as it
does in the Weyl-Kondo model (Sect.\,\ref{WKSM}). The Weyl node 1  with sizeable
tilt might be a good candidate for this process. Whereas in previous DFT +
dynamical mean field theory (DMFT) calculations for Ce$_3$Bi$_4$Pd$_3$
\cite{Tom18.1} topologically nontrivial band crossings were not considered, a
very recent study \cite{Cao20.2} suggests that nodal lines are also present
in the uncorrelated (DFT) bandstructure, far above the Fermi level. The
low-temperature specific heat of Ce$_3$Bi$_4$Pd$_3$ \cite{Dzs17.1}, however, is
consistent with  a scenario where Weyl point (as opposed to line) nodes are
driven by the Kondo effect to the immediate vicinity  of the Fermi energy (as
described by our Weyl-Kondo model, Sect.\,\ref{WKSM}).

\section{Theoretical treatment of tilted Weyl-Kondo semimetal model}\label{SImodel}

\subsection{Microscopic model}\label{WKSM}

\noindent To study the effect of the Kondo interaction on a conduction electron
band with tilted Weyl cones, we generalize a previously established Weyl-Kondo
semimetal model \cite{Lai18.1} (for which the role of nonsymmorphic space-group symmetry was recently emphasized \cite{Gre20.1}) to the case
with tilting. The periodic Anderson model reads \vspace{-0.8cm}

\begin{eqnarray}
H & = & H_c + H_{cd} + H_d \;\; , \; \mbox{with} \label{PAM1} \\
& & H_c=\sum_{\bm{k}}\Psi^\dagger_{\bm{k}}~h_{\bm{k}}~\Psi_{\bm{k}} \;\; , \; \mbox{where}\label{eq:hc_sigma}\\
& & \quad\quad \Psi^T_{\bm{k}}=\begin{pmatrix} c_{{\bm{k}}\up,A}& c_{{\bm{k}}\up,B} & c_{{\bm{k}}\dn,A} & c_{{\bm{k}}\dn,B}\end{pmatrix} \;\; , \; \mbox{and}\nonumber \\
& & \quad\quad h_{\bm{k}}= \sigma_0[ u_1({\bm{k}})\tau_x + u_2({\bm{k}})\tau_y + m\tau_z] + \lambda [ {\bm{D}({\bm{k}})} \cdot \bm{ \sigma}] \tau_z + C
[ m - \lambda D({\bm{k}}) ]
\sigma_0 \tau_0 \;\; , \quad\quad \label{eq:hc_sigma}\nonumber \\
& & H_{cd} =
V \sum_{i,\sigma} \left( d_{i \sigma}^\dagger c_{i \sigma} + \Hc\right) 
\;\; , \; \mbox{and} \label{eq:hybridization}\\
& & H_d=E_{d} \sum_{i, \sigma} d_{i \sigma}^\dagger d_{i \sigma} + U \sum_i n^d_{i \up} n^d_{i \dn} \;\; . \label{eq:hd}
\end{eqnarray}

\noindent Here, $H_c$, $H_{cd}$, and $H_d$ are the conduction electron term,
hybridization term, and strongly correlated $d$ electron term (representing the
physical $4f$ electrons), respectively. The notations are the same as in
ref\,\cite{Lai18.1}. In particular, $u_1({\bm{k}})$ and $u_2({\bm{k}})$ are
determined by the conduction electron hopping, $t_{\la ij \ra } = t$ between
nearest-neighbor sites ($\la ij \ra $). The second term specifies a
Dresselhaus-type spin-orbit coupling between the second-nearest-neighbor sites
($\la \la ij \ra \ra$), which is of strength $\lambda$ and involves vector
${\bm{D}}({\bm{k}})=\begin{pmatrix}D_x({\bm{k}}),D_y({\bm{k}}),D_z({\bf
k})\end{pmatrix}$. Specifically, \vspace{-0.8cm}

\begin{eqnarray}
u_1({\bm{k}}) & = & t\left(1+\sum_{n=1}^3\cos({\bm{k}}\cdot \bm{a}_n)\right) \;\; , \label{eq:d1}\\
u_2({\bm{k}}) & = & t\sum_{n=1}^3 \sin({\bm{k}}\cdot\bm{a}_n ) \;\; , \label{eq:d2}\\
{D_x({\bm{k}})} & = & \sin({\bm{k}}\cdot \bm{a}_2 ) - \sin({\bm{k}}\cdot \bm{a}_3) - \sin({\bm{k}}\cdot (\bm{a}_2 - \bm{a}_1 )) + \sin({\bm{k}}\cdot (\bm{a}_3 -\bm{a}_1)) \;\; , \label{eq:Dx}
\end{eqnarray}

\noindent and $D_y,~D_z$ are obtained by permuting the fcc primitive lattice
vectors $\bm{a}_{n=1,2,3}$. Additionally, $\bm{\sigma} =
(\sigma_x,\sigma_y,\sigma_z)$ and $\bm{\tau} =(\tau_x,\tau_y,\tau_z)$ are the
Pauli matrices acting on the spin and sublattice spaces, respectively, and
$\sigma_0$ is the identity matrix. The last term in $H_c$, $C [m - \lambda
D(\bm{k}) ] \sigma_0 \tau_0$, is the tilt (or ``$C$'') term and has not been
considered before. Here, $D(\bm{k}) \equiv \left| \bm{D}(\bm{k}) \right| =
\sqrt{ D_x(\bm{k})^2 + D_y(\bm{k})^2 + D_z(\bm{k})^2}$. Physically, the $C$ term
represents kinetic hopping that goes beyond the nearest-neighboring sites.

By solving the self-consistent saddle-point equations for the strong interaction
limit ($U=\infty$), we find a Weyl-Kondo solution in the presence of the $C$
term. Specifically, we follow the approach in ref\,\cite{Lai18.1} to take care
of the prohibition of $d$ fermion double occupancy by an auxiliary-particle
method \cite{Hew97.1}, which rewrites $d^\dagger_{i\sigma} = f^\dagger_{i\sigma}
b_{i}$. The $f^\dagger_{i\sigma}$ ($b_i$) are fermionic (bosonic) operators,
which satisfy a constraint that is enforced by a Lagrange multiplier $\ell$.
This approach leads to a set of saddle-point equations, where $b_i$ condenses to
a value $r$. The $d$ fermion level and the hybridization get renormalized as
$E_d \rightarrow E_d + \ell$ and $ V\rightarrow rV$.
Figure~\ref{Fig:Dispersion_TWKSM} describes the renormalized bands, and
illustrates the tilting of the Weyl-Kondo cone at the Fermi energy.

\begin{figure}[t!] 
\includegraphics[width=0.55\textwidth]{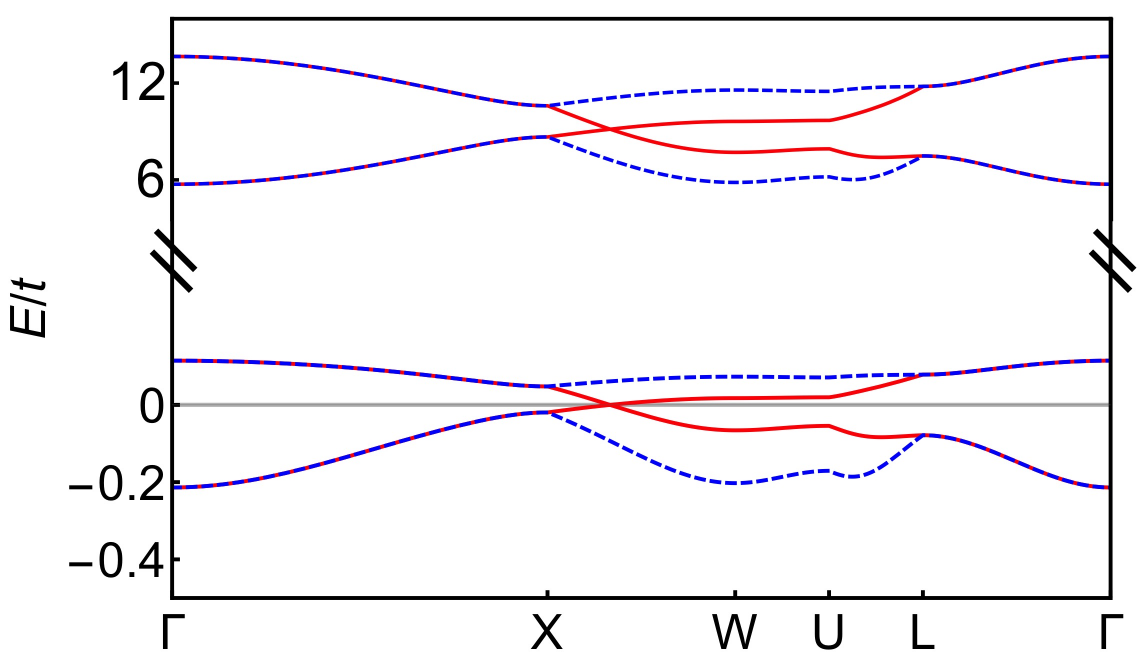} 

\caption{\label{Fig:Dispersion_TWKSM} {\bf Kondo interaction pins tilted Weyl
nodes to Fermi energy.}  Dispersion for the tilted Weyl-Kondo semimetal model of
Eqns.\,\ref{PAM1}-\ref{eq:hd}, with energy in units of the hopping
amplitude $t$, the Fermi energy at $E/t = 0$, and the parameters $(t, \lambda,
m, E_d, V, C) = (1, 0.5, 1, -6, 6.6, 0.9)$; in the resulting saddle-point
solution, $ r \simeq 0.2614$ and $\ell \simeq 6.3366$. The nonzero $C$ term
causes the Weyl nodes to be tilted, which is seen in the dispersion near the
Weyl nodes between X and W. The upper quartet of bands, with the bandwidth
$D/t$ of the conduction electrons, is a remnant of the solution in the absence of the Kondo effect. That it contains Weyl nodes far away from the Fermi level captures the conduction-electron only solution of our open-core DFT calculations (Fig.\,\ref{Fig4}A). The
lower quartet of bands, with strongly renormalized bandwidth ($k_{\rm B}T_{\rm
K}/t$, where $T_{\rm K}$ is the Kondo temperature), appears only in the presence
of the Kondo interaction. It contains \qs{emergent Weyl nodes at}
the Fermi energy. Away from the commensurate filling used here, they will be
situated slightly away from the Fermi level, as sketched in Fig.\,\ref{Fig4}B.}
\end{figure}

\subsection{Berry curvature induced Hall response of titled Weyl-Kondo semimetal: Perturbative regime}\label{WKSMlow}

\noindent To illustrate the role of the tilting term in the Weyl-Kondo
semimetal, we first consider the weak-field limit.  Here, it was shown
\cite{Sod15.1} that the Hall conductivity $\sigma_{xy}$ contains only $2\omega$
(and the associated $0\omega$) components and is determined by the Berry
curvature dipole $\bm{D}$, a rank-2 tensor with components $D_{ab=x,y,z}$ that
are expressed as \vspace{-0.8cm}

\begin{eqnarray}
D_{ab} & = & \int  {\bar{d}}^{\,3}k ~ f_0(\varepsilon_{\bm{k}})  ~\partial_{ka}
\Omega_b(\bm{k}) \label{j2omega2} \\
& = &\oint_{\rm{FS}}  {\bar{d}}^{\,2}k ~ \frac{v_a (\bm{k}) \Omega_b(\bm{k})}{v(\bm{k})} \quad , 
\label{Eq:BCD_FS2}
\end{eqnarray}

\noindent where ${\bar{d}}^n k \equiv d^n k/ (2\pi)^n$, $\partial_{k a} \equiv
\partial/\partial_{k a}$, $v_a(\bm{k}_{\rm{F}})= \partial_{k a} \varepsilon_{k
a}$ is the velocity, and $f_0(\varepsilon_{\bm{k}})$ is the equilibrium
(Fermi-Dirac) distribution function of eigenenergy $\varepsilon_{\bm{k}}$, with
the Fermi energy being set to $0$; at $T\rightarrow0$, 
$f_0(\varepsilon_{\bm{k}}) = \Theta(-\varepsilon_{\bm{k}})$. We consider the 
Fermi pockets to surround the Weyl nodes, and have used the Jacobian, $|\nabla
\varepsilon_{\bm{k}}| = \sqrt{\sum_{a} v_a (\bm{k})^2} \equiv v(\bm{k})$, in
converting the bulk integral into the integration
$\oint_{\rm{FS}}{\bar{d}}^{\,2} k$  on the Fermi surface.

We consider a Weyl ($\rm{W}^+$) and an anti-Weyl ($\rm{W}^-$) node, located on
the $k_x$-$k_y$ plane sketched in Fig.\,\ref{Fig:tilted_nodes}A (for the 3D
Brillouin zone, see Fig.\,\ref{BZ}). The considered zincblende structure has a
mirror symmetry $M_{xy}$, indicated by the dashed $k_x=k_y$ line in
Fig.\,\ref{Fig:tilted_nodes}A. The dispersion of the Weyl-Kondo solution across
$\rm{W}^+$ and $\rm{W}^-$, perpendicular to this line, is shown in
Fig.\,\ref{Fig:tilted_nodes}B for the case where the Fermi energy is slightly
away from the nodal energy (a close-up in energy of this dispersion is shown in
Fig.\,\ref{Fig4}B of the main part). $M_{xy}$ reflects the Fermi pocket around
$\rm{W}^+$ onto that of $\rm{W}^-$ (see black Fermi surface contour lines in
Fig.\,\ref{Fig:tilted_nodes}A). Under $M_{xy}$, $\bm{k}$ and $\bm{v}$ transform
as vectors, while $\bf{\Omega}$ transforms as a pseudovector. Each component
transforms as $k_{\rm{F},x/y}^{\rm{W}-} \leftrightarrow
k_{\rm{F},y/x}^{\rm{W}+}$, $v_{x/y}^{\rm{W}-} \leftrightarrow
v_{y/x}^{\rm{W}+}$, $\Omega_{x/y}^{\rm{W}-}\leftrightarrow
-\Omega_{y/x}^{\rm{W}+}$. We find that, under $M_{xy}$, the contribution from
the Fermi surface around $\rm{W}^+$ and that from the Fermi surface around
$\rm{W}^-$ are related to each other as $D_{xy}^{\rm{W}+}(\bm{k}_{\rm
F}^{\rm{W}+}) \leftrightarrow -D_{yx}^{\rm{W}-}(\bm{k}_{\rm F}^{\rm{W}-})$. We
can thus re-express $D_{xy}$ as \vspace{-0.4cm}

\begin{equation}
\hspace{-0.2cm} D_{xy} = D_{xy}^{\rm{W}+}(\bm{k}_{\rm F}^{\rm{W}+}) + D_{xy}^{\rm{W}-}(\bm{k}_{\rm F}^{\rm{W}-}) =
D_{xy}^{\rm{W}+}(\bm{k}_{\rm F}^{\rm{W}+}) - D_{yx}^{\rm{W}+}(\bm{k}_{\rm F}^{\rm{W}+}) = \oint_{\rm{FS}}^{\rm{W}+} \bar{d}^{\,2}
k ~~\bm{\hat{z}} \cdot \left(\bm{\hat{v}} \times \bm{\Omega}\right) \;\; ,
\label{Eq:Dxy_prod_form}
\end{equation}
\noindent where we have used Eqn.\,\ref{Eq:BCD_FS2}. The integral $\oint$ is now
only on the Fermi surface around the $\rm{W}^+$ node. Here, $\bm{\hat{v}}$
represents the unit vector along the direction of the Fermi velocity vector, 
and $\bm{\Omega} = \begin{pmatrix} \Omega_x & \Omega_y & \Omega_z
\end{pmatrix}$. We note that, in addition to the ones of the $k_z=2\pi$ plane we
have discussed so far, Weyl/anti-Weyl nodes also exist on the planes $k_x =
2\pi$ and $k_y = 2\pi$. The $M_{xy}$ operation relates the Weyl nodes on the
$k_x = 2\pi$ plane to the anti-Weyl nodes on the $k_y = 2\pi$ plane. Thus, all
these nodes contribute to the Berry curvature dipole according to
Eqn.\,\ref{Eq:Dxy_prod_form}.

\begin{figure}[t!] 
\includegraphics[width=0.29\textwidth]{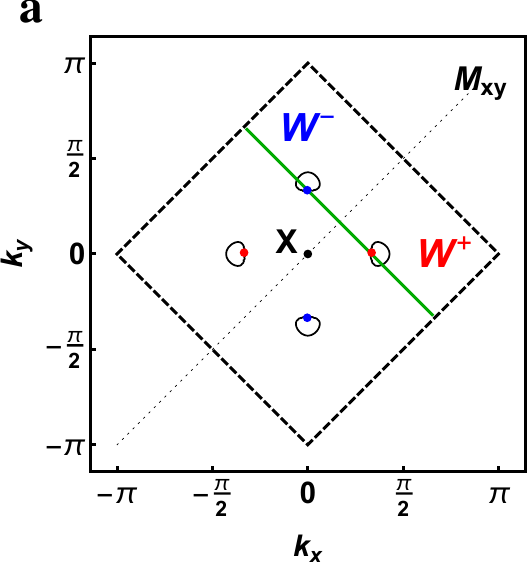} \hspace*{0.1cm}
\includegraphics[width=0.3\textwidth]{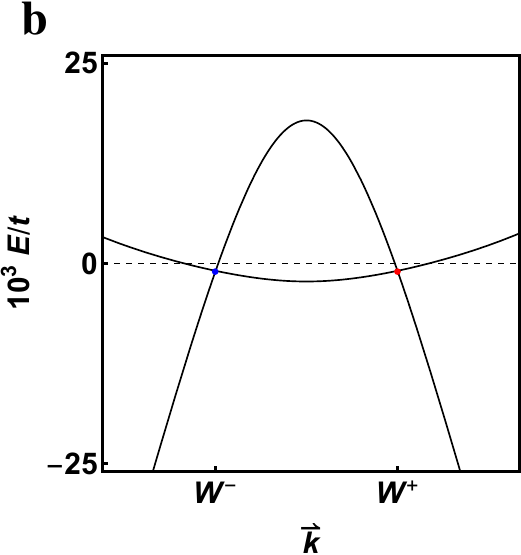} \hspace*{0.1cm} \includegraphics[width=0.3\textwidth]{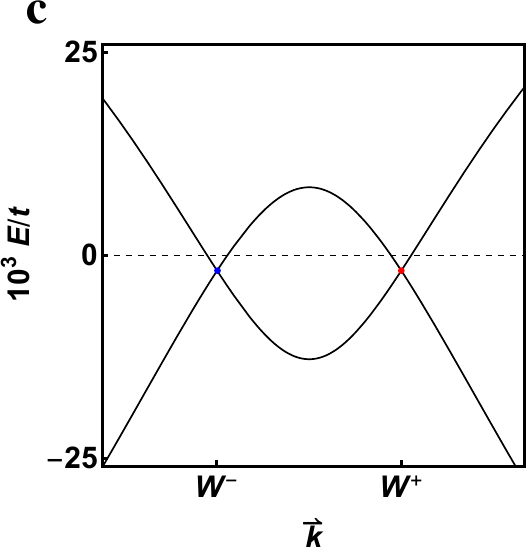}

\caption{\label{Fig:tilted_nodes} {\bf Fermi pockets and dispersion around pairs
of Weyl nodes.} ({\bf A}) $k_x$-$k_y$ plane at the Brillouin zone boundary $k_z =
2\pi$, with two pairs of Weyl nodes as obtained for the tilted Weyl-Kondo
semimetal model of Eqns.\,\ref{PAM1}-\ref{eq:hd}, for the tilt parameter $C=0.9$
and the Fermi energy slightly away from the nodal energy. The green line
indicates a path through the Weyl node $\rm{W}^+$ and the anti-Weyl node
$\rm{W}^-$ perpendicular to the dotted $k_x=k_y$ line representing the $M_{xy}$
mirror symmetry of the considered zincblende structure. The black contours are
the 2D projections of Fermi pockets surrounding the nodes. ({\bf B}) Dispersion
along the green line in ({\bf A}), with energy in units of $t$, the Fermi energy
at $E/t=0$, and for $C=0.9$. ({\bf C}) Corresponding dispersion without tilt
($C=0$), as determined in ref\,\cite{Lai18.1}.}
\end{figure}

For a Weyl node without tilt, the Fermi surface is a sphere with the node in its
center. The unit Fermi velocity $\bm{\hat{v}}$ in the integral in
Eqn.\,\ref{Eq:Dxy_prod_form} is then always parallel or antiparallel to the
Berry curvature vector $\bm{\Omega}$, which results in $\bm{\hat{v}} \times
\bm{\Omega} = 0$. In the presence of tilt, by contrast, $\bm{\hat{v}} \times
\bm{\Omega} \neq 0$ at any momentum point on the Fermi surface because
$\bm{\hat{v}}$ is not (anti)parallel to $\bm{\Omega}$. Moreover, because the
Kondo effect places the Weyl/anti-Weyl nodes close to the Fermi energy, the
Berry curvature $\bm{\Omega}$ is very large on the Fermi surface, leading thus
to very large values of $D_{xy}$.

It is straightforward to generalize Eqn.\,\ref{Eq:Dxy_prod_form} for arbitrary
directions. For the $D_{yz}$ and $D_{xz}$ components, we can utilize $M_{yz}$
($M_{xz}$) to arrive at similar results with $\bm{\hat{z}} \rightarrow
\bm{\hat{x}}~(\bm{\hat{y}})$. In general, $D_{ab}$ can be written in a compact
form as \vspace{-0.4cm}

\begin{equation}
D_{ab} = \oint_{\rm{FS}}^{\rm{W}+} {\bar{d}}^{\,2}k ~~ \bm{\hat{c}}\cdot \left( \bm{\hat{v}} \times \bm{\Omega} \right) \;\; ,
\end{equation}
\noindent where $\bm{\hat{c}}$ refers to the axis perpendicular to $a$ and $b$.

Our experiments, however, have revealed a spontaneous Hall effect also in the
first harmonic response and can thus not be explained by this weak-field
treatment. In Sect.\,\ref{SINLH} below we show that, in the fully nonequilibrium
regime, both the second and first harmonic response are allowed.

\subsection{Berry curvature induced Hall response of titled Weyl-Kondo
semimetal: Fully nonequilibrium regime}\label{SINLH}

\noindent Weyl nodes are Berry curvature singularities. In the Weyl-Kondo
semimetal phase, these nodes are pinned close to the Fermi energy
\cite{Lai18.1}. This has two immediate consequences. First, the Berry curvature
singularities dominate the electronic behavior. Second, the Fermi pockets are
very small. In addition, for tilted Weyl cones, the Fermi surface is asymmetric
with respect to the Weyl nodes and there will thus be directions along which the
Weyl nodes are in extreme proximity (at a minimal distance denoted by $k_{\rm
W}$) to the Fermi surface, as is sketched in Fig.\,\ref{Fig4}C of the main part.
An applied electric field will then readily cause a shift $\Delta\bm{k}$ in the
distribution function $f(\bm{k})$ that is sizeable compared to $k_{\rm W}$, and
possibly even compared to the Fermi wavevector $k_{\rm F}$. This brings the
system to the fully nonequilibrium regime (Fig.\,\ref{BZ}), where the
perturbative treatment of Sect.\,\ref{WKSMlow} will fail.

In fact, an electric field $\bm{\mathcal{E}}=\mathcal{E}_x(t)\bm{\hat{x}}$ has a
current response not only along the longitudinal $x$ direction but, via
\cite{Sun99.1} \vspace{-0.8cm}

\begin{eqnarray}
v_{{\rm{an}},y}(\bm{k}) & = & -\frac{e}{\hbar} \mathcal{E}_x(t)\Omega_z(\bm{k}) \;\; , \label{Delta-vy}\\
v_{{\rm{an}},z}(\bm{k}) & = & \frac{e}{\hbar} \mathcal{E}_x(t)\Omega_y(\bm{k})\;\; , \label{Delta-vz}
\end{eqnarray}

\noindent also along the transverse  directions. In the Hall effect geometry,
the ensuing transverse currents will set up transverse electric fields
$\mathcal{E}_y$ and $\mathcal{E}_z$. The total electric field will produce a
distribution function \vspace{-0.4cm}

\begin{equation}
f(\bm{k})=f_0(\bm{k})+g(\bm{k}) \label{totalf}\;\; ,
\end{equation}
\noindent that will be appreciably different from the equilibrium (Fermi-Dirac)
distribution function $f_0(\bm{k})$, and even break the lattice symmetry
(Fig.\,\ref{BZ}). Thus, an expansion around $f_0(\bm{k})$ will not be
sufficient.

\begin{figure}[t!]
\begin{center}
\includegraphics[width=0.4\textwidth]{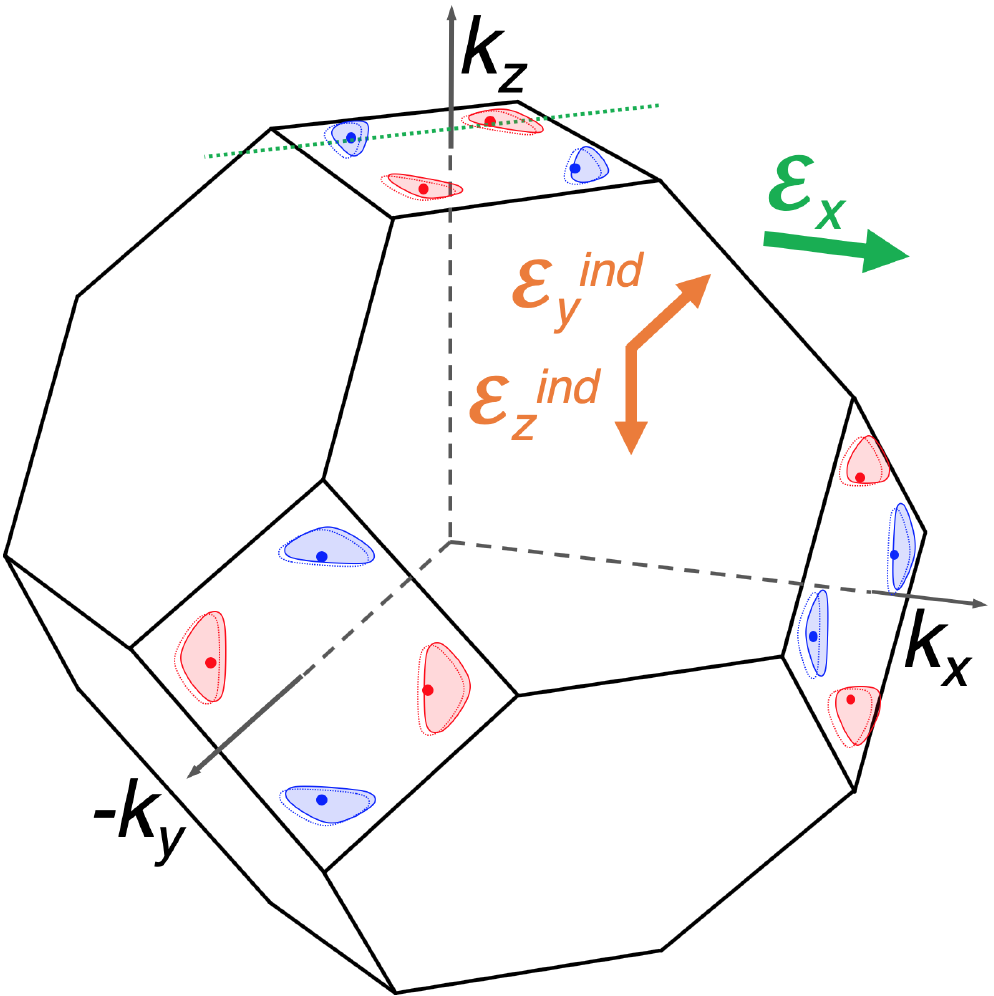}
\end{center}

\caption{\label{BZ} {\bf Electric field effect on the electron distribution
in the fully nonequilibrium regime.} Schematic of the Brillouin zone (BZ) of the
tilted Weyl-Kondo semimetal model of Sect.\,\ref{WKSM}, with the Weyl nodes
shown as dots and the Fermi pockets projected to the BZ boundary indicated by
red and blue dotted lines. The green dotted line indicates the cut shown in
Fig.\,\ref{Fig4}B of the main part. An electric field $\mathcal{E}_x$ applied
along the $k_x$ direction induces, via the anomalous velocity $\bm{v}_{\rm{an}}
\sim \bm{\mathcal{E}}\times\bm{\Omega}$, also transverse fields $\mathcal{E}_y$
and $\mathcal{E}_z$. The proximity of the Weyl and anti-Weyl nodes to the Fermi
surface implies that  even an electric field small in absolute scale causes a
nonperturbative change to the  electron distribution, shifting its occupied part
in all directions (red and blue shaded areas).}
\end{figure}

To delineate the relation to previous (perturbative) approaches \cite{Sod15.1},
we distinguish a weak-field regime, where $\Delta k/k_{\rm{F}}$ and $\Delta
k/k_{\rm{W}}$ are both $\ll 1$ (Fig.\,\ref{Fig4}C\,left of the main part), and a
fully nonequilibrium regime where either of the two (or both) are no longer true
(Fig.\,\ref{Fig4}C\,right of the main part and Fig.\,\ref{BZ}). The reference
point for the weak-field regime is $\mathcal{E}=0$ (Fig.\,\ref{Fig4}D of the
main part), where the system is in equilibrium and the corresponding
distribution function is $f_0(\bm{k})$, which is even in $\bm{k}$. Replacing
$f(\bm{k})$ in \vspace{-0.4cm}

\begin{equation} 
\sigma_{xy} = \frac{e^2}{\hbar}\int_{\bm{k}} f(\bm{k})\Omega_z(\bm{k})
\label{eq:1omega}
\end{equation}
\noindent by $f_0(\bm{k})$ (the zeroth order term of the Taylor expansion) leads
to a vanishing result as $\Omega_z(\bm{k})$ is odd in $\bm{k}$. The first order
term in the Taylor expansion, which is linear in $\mathcal{E}$, leads to a
linear-in-$\mathcal{E}$ Hall conductivity $\sigma_{xy}$ and, via\vspace{-0.4cm}

\begin{equation}
j_y = \sigma_{xy}(\bm{\mathcal{E}})\cdot\mathcal{E}_x \;\; , \label{eq:jy}
\end{equation}
\noindent to a linear-in-$\mathcal{E}^2$ Hall current density $j_y$ that
corresponds to the 2$\omega$ contribution to $\sigma_{xy}$ derived in ref\,\cite{Sod15.1}.

By contrast, in the fully nonequilibrium regime (Fig.\,\ref{Fig4}C right and
\ref{Fig4}D of the main part and Fig.\,\ref{BZ}), $\Delta{k}/k_{\rm W}$ (and
possibly $\Delta{k}/k_{\rm F}$ as well) not being small means that we can no
longer expand $f(\bm{k})$ with respect to the equilibrium distribution. Instead,
the reference point corresponds to a nonzero drive $\mathcal{E}$, where the
distribution function $f(\bm{k})$ contains an odd-in-$\bm{k}$ component.
Inserting such an $f(\bm{k})$ term into Eqn.\,\ref{eq:1omega} leads to a nonzero
value for the zeroth order term in $\sigma_{xy}$, and thus to a $1\omega$
response as observed in our experiments. Accompanying it will be a $2\omega$
(and an associated $0\omega$) response, that comes from the first-order term in
the Taylor expansion of $f(\bm{k})$ with respect to the nonzero $\mathcal{E}$.
Of course, a Taylor expansion around a nonzero field is again an approximation
and, ultimately, $f(\bm{k})$ should be determined in a fully nonperturbative
approach, which is a challenging task for future studies.

In summary, the appearance of the $1\omega$ component of the spontaneous Hall
response in the fully nonequilibrium regime is a new effect; it is distinct from
the weak-field regime \cite{Sod15.1}, where this component is absent. As
stressed above, a Weyl-Kondo semimetal is readily---without applying electric
fields that are large on absolute scales---in this regime due to its small
Fermi pockets that asymmetrically surround the Weyl and anti-Weyl nodes.


\begin{thebibliography}{10}

\bibitem{NatPhys16.1a}
{Focus Issue ``Topological matter''}, {{\em Nat.\ Phys.} {\bf 12}, 615 (2016)}.

\bibitem{Koe07.1}
M.~K\"onig, S.~Wiedmann, C.~Br\"une, A.~Roth, H.~Buhmann, L.~W. Molenkamp,
  X.~Qi, S.~Zhang, {Quantum spin Hall insulator state in HgTe quantum wells}.
\newblock {\it Science\/} {\bf 318}, 766 (2007).

\bibitem{Hsi09.2}
D.~Hsieh, Y.~Xia, D.~Qian, L.~Wray, J.~H. Dil, F.~Meier, J.~Osterwalder,
  L.~Patthey, J.~G. Checkelsky, N.~P. Ong, A.~V. Fedorov, H.~Lin, A.~Bansil,
  D.~Grauer, Y.~S. Hor, R.~J. Cava, M.~Z. Hasan, {A tunable topological
  insulator in the spin helical Dirac transport regime}.
\newblock {\it Nature\/} {\bf 460}, 1101 (2009).

\bibitem{Sas11.1}
S.~Sasaki, M.~Kriener, K.~Segawa, K.~Yada, Y.~Tanaka, M.~Sato, Y.~Ando,
  {Topological superconductivity in Cu$_x$Bi$_2$Se$_3$}.
\newblock {\it {Phys.\ Rev.\ Lett.}\/} {\bf 107}, 217001 (2011).

\bibitem{Wu16.1}
L.~Wu, S.~Patankar, T.~Morimoto, N.~L. Nair, E.~Thewalt, A.~Little, J.~G.
  Analytis, J.~E. Moore, J.~Orenstein, {Giant anisotropic nonlinear optical
  response in transition metal monopnictide Weyl semimetals}.
\newblock {\it {Nat.\ Phys.}\/} {\bf 13}, 350 (2016).

\bibitem{Ma17.1}
Q.~Ma, S.-Y. Xu, C.-K. Chan, C.-L. Zhang, G.~Chang, Y.~Lin, W.~Xie,
  T.~Palacios, H.~Lin, S.~Jia, P.~A. Lee, P.~Jarillo-Herrero, N.~Gedik, {Direct
  optical detection of Weyl fermion chirality in a topological semimetal}.
\newblock {\it {Nat.\ Phys.}\/} {\bf 13}, 842 (2017).

\bibitem{Arm18.1}
N.~P. Armitage, E.~J. Mele, A.~Vishwanath, {Weyl and Dirac semimetals in
  three-dimensional solids}.
\newblock {\it {Rev.\ Mod.\ Phys.}\/} {\bf 90}, 015001 (2018).

\bibitem{Hua15.1}
S.-M. Huang, S.-Y. Xu, I.~Belopolski, C.-C. Lee, G.~Chang, B.~Wang,
  N.~Alidoust, G.~Bian, M.~Neupane, C.~Zhang, S.~Jia, A.~Bansil, H.~Lin, M.~Z.
  Hasan, {A Weyl fermion semimetal with surface Fermi arcs in the transition
  metal monopnictide TaAs class}.
\newblock {\it {Nat.\ Commun.}\/} {\bf 6}, 7373 (2015).

\bibitem{Xu15.2}
S.-Y. Xu, I.~Belopolski, N.~Alidoust, M.~Neupane, G.~Bian, C.~Zhang, R.~Sankar,
  G.~Chang, Z.~Yuan, C.-C. Lee, S.-M. Huang, H.~Zheng, J.~Ma, D.~S. Sanchez,
  B.~Wang, A.~Bansil, F.~Chou, P.~P. Shibayev, H.~Lin, S.~Jia, M.~Z. Hasan,
  {Discovery of a Weyl fermion semimetal and topological Fermi arcs}.
\newblock {\it {Science}\/} {\bf 349}, 613 (2015).

\bibitem{Xu16.3}
N.~Xu, H.~M. Weng, B.~Q. Lv, C.~E. Matt, J.~Park, F.~Bisti, V.~N. Strocov,
  D.~Gawryluk, E.~Pomjakushina, K.~Conder, N.~C. Plumb, M.~Radovic, G.~Aut{\`e}s,
  O.~V. Yazyev, Z.~Fang, X.~Dai, T.~Qian, J.~Mesot, H.~Ding, M.~Shi,
  {Observation of Weyl nodes and Fermi arcs in tantalum phosphide}.
\newblock {\it {Nat.\ Commun.}\/} {\bf 7}, 11006 (2016).

\bibitem{Hua15.2}
X.~Huang, L.~Zhao, Y.~Long, P.~Wang, D.~Chen, Z.~Yang, H.~Liang, M.~Xue,
  H.~Weng, Z.~Fang, X.~Dai, G.~Chen, {Observation of the chiral-anomaly-induced
  negative magnetoresistance in 3D Weyl semimetal TaAs}.
\newblock {\it {Phys.\ Rev.\ X}\/} {\bf 5}, 031023 (2015).

\bibitem{Zha16.3}
C.-L. Zhang, S.-Y. Xu, I.~Belopolski, Z.~Yuan, Z.~Lin, B.~Tong, G.~Bian,
  N.~Alidoust, C.-C. Lee, S.-M. Huang, T.-R. Chang, G.~Chang, C.-H. Hsu, H.-T.
  Jeng, M.~Neupane, D.~S. Sanchez, H.~Zheng, J.~Wang, H.~Lin, C.~Zhang, H.-Z.
  Lu, S.-Q. Shen, T.~Neupert, M.~Zahid~Hasan, S.~Jia, {Signatures of the
  Adler--Bell--Jackiw chiral anomaly in a Weyl fermion semimetal}.
\newblock {\it {Nat.\ Commun.}\/} {\bf 7}, 10735 (2016).

\bibitem{Mol16.1}
P.~J.~W. Moll, N.~L. Nair, T.~Helm, A.~C. Potter, I.~Kimchi, A.~Vishwanath,
  J.~G. Analytis, {Transport evidence for Fermi-arc-mediated chirality transfer
  in the Dirac semimetal Cd$_3$As$_2$}.
\newblock {\it Nature\/} {\bf 535}, 266 (2016).

\bibitem{Zha19.2}
C.~Zhang, Y.~Zhang, X.~Yuan, S.~Lu, J.~Zhang, A.~Narayan, Y.~Liu, H.~Zhang,
  Z.~Ni, R.~Liu, E.~S. Choi, A.~Suslov, S.~Sanvito, L.~Pi, H.-Z. Lu, A.~C.
  Potter, F.~Xiu, {Quantum Hall effect based on Weyl orbits in Cd$_3$As$_2$}.
\newblock {\it {Nature}\/} {\bf 565}, 331 (2019).

\bibitem{Han19.1}
S.~Han, C.~Lee, E.-G. Moon, H.~Min, {Emergent anisotropic non-Fermi liquid at a
  topological phase transition in three dimensions}.
\newblock {\it Phys. Rev. Lett.\/} {\bf 122}, 187601 (2019).

\bibitem{Men19.2}
T.~Meng, J.~C. Budich, {Unpaired Weyl nodes from long-ranged interactions: fate
  of quantum anomalies}.
\newblock {\it Phys. Rev. Lett.\/} {\bf 122}, 046402 (2019).

\bibitem{Kan20.2}
M.~Kang, L.~Ye, S.~Fang, J.-S. You, A.~Levitan, M.~Han, J.~I. Facio,
  C.~Jozwiak, A.~Bostwick, E.~Rotenberg, M.~K. Chan, R.~D. McDonald, D.~Graf,
  K.~Kaznatcheev, E.~Vescovo, D.~C. Bell, E.~Kaxiras, J.~van~den Brink,
  M.~Richter, M.~Prasad~Ghimire, J.~G. Checkelsky, R.~Comin, {Dirac fermions
  and flat bands in the ideal kagome metal FeSn}.
\newblock {\it {Nat.\ Mater.}\/} {\bf 19}, 163 (2020).

\bibitem{Sha20.1}
Y.~Shao, A.~N. Rudenko, J.~Hu, Z.~Sun, Y.~Zhu, S.~Moon, A.~J. Millis, S.~Yuan,
  A.~I. Lichtenstein, D.~Smirnov, Z.~Q. Mao, M.~I. Katsnelson, D.~N. Basov,
  {Electronic correlations in nodal-line semimetals}.
\newblock {\it {Nat.\ Phys.}\/} {\bf 16}, 636 (2020).

\bibitem{Yan14.2}
B.-J. Yang, E.-G. Moon, H.~Isobe, N.~Nagaosa, {Quantum criticality of
  topological phase transitions in three-dimensional interacting electronic
  systems}.
\newblock {\it {Nat.\ Phys.}\/} {\bf 10}, 774 (2014).

\bibitem{Par16.1}
W.~K. Park, L.~Sun, A.~Noddings, D.-J. Kim, Z.~Fisk, L.~H. Greene, {Topological
  surface states interacting with bulk excitations in the Kondo insulator
  SmB$_6$ revealed via planar tunneling spectroscopy}.
\newblock {\it {Proc.\ Natl.\ Acad.\ Sci.\ U.S.A.}\/} {\bf 113}, 6599 (2016).

\bibitem{Cas17.1}
D.~Castelvecchi, {The shape of things to come}.
\newblock {\it {Nature}\/} {\bf 547}, 272 (2017).

\bibitem{Ipp18.1}
M.~Ippoliti, R.~N. Bhatt, F.~D.~M. Haldane, {Geometry of flux attachment in
  anisotropic fractional quantum Hall states}.
\newblock {\it Phys. Rev. B\/} {\bf 98}, 085101 (2018).

\bibitem{Rah19.1}
A.~Rahmani, M.~Franz, {Interacting Majorana fermions}.
\newblock {\it {Rep.\ Prog.\ Phys.}\/} {\bf 82}, 084501 (2019).

\bibitem{Dzs17.1}
S.~Dzsaber, L.~Prochaska, A.~Sidorenko, G.~Eguchi, R.~Svagera, M.~Waas,
  A.~Prokofiev, Q.~Si, S.~Paschen, {Kondo insulator to semimetal transformation
  tuned by spin-orbit coupling}.
\newblock {\it {Phys.\ Rev.\ Lett.}\/} {\bf 118}, 246601 (2017).

\bibitem{Lai18.1}
H.-H. Lai, S.~E. Grefe, S.~Paschen, Q.~Si, {Weyl-Kondo semimetal in
  heavy-fermion systems}.
\newblock {\it {Proc.\ Natl.\ Acad.\ Sci.\ U.S.A.}\/} {\bf 115}, 93 (2018).

\bibitem{Sin14.2}
Y.~P. Singh, D.~J. Haney, X.~Y. Huang, I.~K. Lum, B.~D. White, M.~Dzero, M.~B.
  Maple, C.~C. Almasan, {From local moment to mixed-valence regime in
  Ce$_{1-x}$Yb$_x$CoIn$_5$ alloys}.
\newblock {\it {Phys.\ Rev.\ B}\/} {\bf 89}, 115106 (2014).

\bibitem{Map06.1}
M.~B. Maple, N.~P. Butch, N.~A. Frederick, P.-C. Ho, J.~R. Jeffries, T.~A.
  Saylesa, T.~Yanagisawa, W.~M. Yuhasz, S.~Chi, H.~J. Kang, J.~W. Lynn, P.~Dai,
  S.~K. McCall, M.~W. McElfresh, M.~J. Fluss, Z.~Henkie, A.~Pietraszko,
  {Field-dependent ordered phases and Kondo phenomena in the filled
  skutterudite compound PrOs$_4$As$_{12}$}.
\newblock {\it {Proc.\ Natl.\ Acad.\ Sci.\ U.S.A.}\/} {\bf 103}, 6783 (2006).

\bibitem{Sch83.1}
P.~Schlottmann, {Bethe-Ansatz solution of the ground-state of the SU $(2j+1)$
  Kondo (Coqblin-Schrieffer) model: Magnetization, magnetoresistance and
  universality}.
\newblock {\it {Z.\ Phys.\ B}\/} {\bf 51}, 223 (1983).

\bibitem{Nag10.1}
N.~Nagaosa, J.~Sinova, S.~Onoda, A.~H. MacDonald, N.~P. Ong, {Anomalous Hall
  effect}.
\newblock {\it {Rev.\ Mod.\ Phys.}\/} {\bf 82}, 1539 (2010).

\bibitem{Liu18.1}
E.~Liu, Y.~Sun, N.~Kumar, L.~Muechler, A.~Sun, L.~Jiao, S.-Y. Yang, D.~Liu,
  A.~Liang, Q.~Xu, J.~Kroder, V.~S\"u{\ss}, H.~Borrmann, C.~Shekhar, Z.~Wang,
  C.~Xi, W.~Wang, W.~Schnelle, S.~Wirth, Y.~Chen, S.~T.~B. Goennenwein,
  C.~Felser, {Giant anomalous Hall effect in a ferromagnetic kagome-lattice
  semimetal}.
\newblock {\it {Nat.\ Phys.}\/} {\bf 14}, {1125} (2018).

\bibitem{Cas45.1}
H.~B.~G. Casimir, {On Onsager's principle of microscopic reversibility}.
\newblock {\it {Rev.\ Mod.\ Phys.}\/} {\bf 17}, 343 (1945).

\bibitem{Sod15.1}
I.~Sodemann, L.~Fu, {Quantum nonlinear Hall effect induced by Berry curvature
  dipole in time-reversal invariant materials}.
\newblock {\it {Phys.\ Rev.\ Lett.}\/} {\bf 115}, 216806 (2015).

\bibitem{Xia10.1}
D.~Xiao, M.-C. Chang, Q.~Niu, {Berry phase effects on electronic properties}.
\newblock {\it {Rev.\ Mod.\ Phys.}\/} {\bf 82}, 1959 (2010).

\bibitem{Du19.1}
Z.~Z. Du, C.~M. Wang, S.~Li, H.-Z. Lu, X.~C. Xie, {Disorder-induced nonlinear
  Hall effect with time-reversal symmetry}.
\newblock {\it {Nat.\ Commun.}\/} {\bf 10}, 3047 (2019).

\bibitem{Zha18.1}
Y.~Zhang, Y.~Sun, B.~Yan, {Berry curvature dipole in Weyl semimetal materials:
  An {\em ab initio} study}.
\newblock {\it {Phys.\ Rev.\ B}\/} {\bf 97}, 041101 (2018).

\bibitem{Ma19.1}
Q.~Ma, S.-Y. Xu, H.~Shen, D.~MacNeill, V.~Fatemi, T.-R. Chang, A.~M.
  Mier~Valdivia, S.~Wu, Z.~Du, C.-H. Hsu, S.~Fang, Q.~D. Gibson, K.~Watanabe,
  T.~Taniguchi, R.~J. Cava, E.~Kaxiras, H.-Z. Lu, H.~Lin, L.~Fu, N.~Gedik,
  P.~Jarillo-Herrero, {Observation of the nonlinear Hall effect under
  time-reversal-symmetric conditions}.
\newblock {\it Nature\/} {\bf 565}, 337 (2019).

\bibitem{Kan19.1}
K.~Kang, T.~Li, E.~Sohn, J.~Shan, K.~F. Mak, {Nonlinear anomalous Hall effect
  in few-layer WTe$_2$}.
\newblock {\it {Nat.\ Mater.}\/} {\bf 18}, 324 (2019).

\bibitem{Nan17.1}
S.~Nandy, G.~Sharma, A.~Taraphder, S.~Tewari, {Chiral anomaly as the origin of
  the planar Hall effect in Weyl semimetals}.
\newblock {\it {Phys.\ Rev.\ Lett.}\/} {\bf 119}, 176804 (2017).

\bibitem{Fer87.1}
A.~Fert, P.~M. Levy, {Theory of the Hall effect in heavy-fermion compounds}.
\newblock {\it {Phys.\ Rev.\ B}\/} {\bf 36}, 1907 (1987).

\bibitem{Hun04.1}
M.~F. Hundley, A.~Malinowski, P.~G. Pagliuso, J.~L. Sarrao, J.~D. Thompson,
  {Anomalous $f$-electron Hall effect in the heavy-fermion system Ce$T$In$_5$
  ($T=\mathrm{Co}$, Ir, or Rh)}.
\newblock {\it Phys. Rev. B\/} {\bf 70}, 035113 (2004).

\bibitem{Pas04.1}
S.~Paschen, T.~L\"uhmann, S.~Wirth, P.~Gegenwart, O.~Trovarelli, C.~Geibel,
  F.~Steglich, P.~Coleman, Q.~Si, {Hall-effect evolution across a heavy-fermion
  quantum critical point}.
\newblock {\it {Nature}\/} {\bf 432}, 881 (2004).

\bibitem{Cus12.1}
J.~Custers, K.~Lorenzer, M.~M\"uller, A.~Prokofiev, A.~Sidorenko, H.~Winkler,
  A.~M. Strydom, Y.~Shimura, T.~Sakakibara, R.~Yu, Q.~Si, S.~Paschen,
  {Destruction of the Kondo effect in the cubic heavy-fermion compound
  Ce$_3$Pd$_{20}$Si$_6$}.
\newblock {\it {Nat.\ Mater.}\/} {\bf 11}, 189 (2012).

\bibitem{Kus19.1}
S.~K. Kushwaha, M.~K. Chan, J.~Park, S.~M. Thomas, E.~D. Bauer, J.~D. Thompson,
  F.~Ronning, P.~F.~S. Rosa, N.~Harrison, {Magnetic field-tuned Fermi liquid in
  a Kondo insulator}.
\newblock {\it {Nat.\ Commun.}\/} {\bf 10}, 5487 (2019).

\bibitem{Ber70.1}
L.~Berger, {Side-jump mechanism for the Hall effect of ferromagnets}.
\newblock {\it Phys. Rev. B\/} {\bf 2}, 4559 (1970).

\bibitem{Egu19.1}
G.~Eguchi, S.~Paschen, {Robust scheme for magnetotransport analysis in
  topological insulators}.
\newblock {\it Phys. Rev. B\/} {\bf 99}, 165128 (2019).

\bibitem{Luk98.1}
G.~M. Luke, Y.~Fudamoto, K.~M. Kojima, M.~I. Larkin, J.~Merrin, B.~Nachumi,
  Y.~J. Uemura, Y.~Maeno, Z.~Q. Mao, Y.~Mori, H.~Nakamura, M.~Sigrist,
  {Time-reversal symmetry-breaking superconductivity in Sr$_2$RuO$_4$}.
\newblock {\it {Nature}\/} {\bf 394}, 558 (1998).

\bibitem{Hil09.1}
A.~D. Hillier, J.~Quintanilla, R.~Cywinski, {Evidence for time-reversal
  symmetry breaking in the noncentrosymmetric superconductor LaNiC$_2$}.
\newblock {\it {Phys.\ Rev.\ Lett.}\/} {\bf 102}, 117007 (2009).

\bibitem{Sha18.1}
T.~Shang, G.~M. Pang, C.~Baines, W.~B. Jiang, W.~Xie, A.~Wang, M.~Medarde,
  E.~Pomjakushina, M.~Shi, J.~Mesot, H.~Q. Yuan, T.~Shiroka, {Nodeless
  superconductivity and time-reversal symmetry breaking in the
  noncentrosymmetric superconductor Re$_{24}$Ti}.
\newblock {\it {Phys.\ Rev.\ B}\/} {\bf 97}, 020502 (2018).

\bibitem{Yao11.1}
A.~Yaouanc, P.~{Dalmas de R\'eotier}, {\it {Muon Spin Rotation, Relaxation, and
  Resonance: Applications to Condensed Matter}\/} (Oxford University Press,
  Oxford, 2011).

\bibitem{Sut12.1}
A.~A. Suter, B.~M. Wojek, {Musrfit: A free platform-independent framework for
  $\mu${SR} data analysis}.
\newblock {\it {Phys.\ Procedia}\/} {\bf 30}, 69 (2012).

\bibitem{Bla18.1}
P.~Blaha, K.~Schwarz, G.~K.~H. Madsen, D.~Kvasnicka, J.~Luitz, R.~Laskowski,
  F.~Tran, L.~Marks, Wien2k, an augmented plane wave plus local orbital program
  for calculating the crystal properties, {TU Vienna, Austria (2018).
  http://www.wien2k.at}.

\bibitem{Per96.1}
J.~P. Perdew, K.~Burke, M.~Ernzerhof, {Generalized gradient approximation made
  simple}.
\newblock {\it {Phys.\ Rev.\ Lett.}\/} {\bf 77}, 3865 (1996).

\bibitem{Her08.2}
W.~Hermes, S.~Linsinger, R.~Mishra, R.~P{\"o}ttgen, {Structure and properties
  of Ce$_3$Pd$_3$Bi$_4$, CePdBi, and CePd$_2$Zn$_3$}.
\newblock {\it {Monatsh.\ Chem.}\/} {\bf 139}, 1143 (2008).

\bibitem{Ahm13.1}
S.~J. Ahmed, J.~Kivinen, B.~Zaporzan, L.~Curiel, S.~Pichardo, O.~Rubel,
  {BerryPI: A software for studying polarization of crystalline solids with
  WIEN2k density functional all-electron package}.
\newblock {\it {Comput.\ Phys.\ Commun.}\/} {\bf 184}, 647 (2013).

\bibitem{Wen15.1}
H.~Weng, C.~Fang, Z.~Fang, B.~A. Bernevig, X.~Dai, {Weyl semimetal phase in
  noncentrosymmetric transition-metal monophosphides}.
\newblock {\it {Phys.\ Rev.\ X}\/} {\bf 5}, 011029 (2015).

\bibitem{Tom18.1}
J.~M. Tomczak, {Thermoelectricity in correlated narrow-gap semiconductors}.
\newblock {\it {J.\ Phys.: Condens. Matter}\/} {\bf 30}, 183001 (2018).

\bibitem{Cao20.2}
C.~Cao, G.-X. Zhi, J.-X. Zhu, {From trivial Kondo insulator Ce$_3$Pt$_3$Bi$_4$
  to topological nodal-line semimetal Ce$_3$Pd$_3$Bi$_4$}.
\newblock {\it {Phys.\ Rev.\ Lett.}\/} {\bf 124}, 166403 (2020).

\bibitem{Gre20.1}
S.~E. Grefe, H.-H. Lai, S.~Paschen, Q.~Si, {Weyl-Kondo semimetals in
  nonsymmorphic systems}.
\newblock {\it Phys. Rev. B\/} {\bf 101}, 075138 (2020).

\bibitem{Hew97.1}
A.~C. Hewson, {\it {The Kondo Problem to Heavy Fermions}\/} (Cambridge
  University Press, Cambridge, 1997).

\bibitem{Sun99.1}
G.~Sundaram, Q.~Niu, {Wave-packet dynamics in slowly perturbed crystals:
  Gradient corrections and Berry-phase effects}.
\newblock {\it {Phys.\ Rev.\ B}\/} {\bf 59}, {14915} (1999).

\end{thebibliography}
\end{document}